\documentclass[revtex4]{emulateapj}

\usepackage{natbib}
\usepackage{graphicx}
\usepackage{longtable}
\usepackage{booktabs}

\newcommand{\kms}{\ifmmode {\rm km~s}^{-1} \else km~s$^{-1}$\fi}
\newcommand{\Msun}{\ifmmode {\rm M}_{\odot} \else $M_{\odot}$\fi }
\newcommand{\Lya}{\ifmmode {\rm Ly}\alpha \else Ly$\alpha$\fi}
\newcommand{\feii}{Fe\,{\sc ii}}
\newcommand{\ciii}{\ifmmode {\rm C}\,{\sc iii}] \else C\,{\sc iii}]\fi}
\newcommand{\civ}{C\,{\sc iv}}
\newcommand{\siiv}{Si\,{\sc iv}}
\newcommand{\aliii}{Al\,{\sc iii}}
\newcommand{\nv}{N\,{\sc v}}
\newcommand{\mgii}{Mg\,{\sc ii}}
\newcommand{\mbh}{$M_{\rm BH}$\ }

\newcommand{\vmin}{$v_{\rm min}$\ } 
\newcommand{\vmax}{$v_{\rm max}$\ }

\shorttitle{BAL Acceleration in SDSS Quasars}
\shortauthors{Grier et al.}

\begin{document}

\title{\civ \ Broad Absorption Line Acceleration in Sloan Digital Sky Survey Quasars}

\author{C.~J.~Grier\altaffilmark{1,2},
W.~N. Brandt\altaffilmark{1,2,3}, 
P.~B.~Hall\altaffilmark{4},
J.~R.~Trump\altaffilmark{1,2,5},
N.~Filiz~Ak\altaffilmark{6}, 
S.~F.~Anderson\altaffilmark{7}, 
Paul~J.~Green\altaffilmark{8}, 
D.~P.~Schneider\altaffilmark{1,2}, 
M.~Sun\altaffilmark{1,9}, 
M.~Vivek\altaffilmark{10}, 
T.~G.~Beatty\altaffilmark{1,11}, 
Joel~R.~Brownstein\altaffilmark{10}, and
Alexandre Roman-Lopes\altaffilmark{12}
}

\altaffiltext{1}{Department of Astronomy \& Astrophysics, The Pennsylvania State University, 525 Davey Laboratory, University Park, PA 16802, USA; grier@psu.edu}
\altaffiltext{2}{Institute for Gravitation and the Cosmos, The Pennsylvania State University, University Park, PA 16802, USA} 
\altaffiltext{3}{Department of Physics, The Pennsylvania State University, University Park, PA 16802, USA}
\altaffiltext{4}{Department of Physics and Astronomy, York University, Toronto, ON M3J 1P3, Canada}
\altaffiltext{5}{Hubble Fellow} 
\altaffiltext{6}{Faculty of Sciences, Department of Astronomy and Space Sciences, Erciyes University, 38039 Kayseri, Turkey}
\altaffiltext{7}{Department of Astronomy, University of Washington, Box 351580, Seattle, WA 98195, USA} 
\altaffiltext{8}{Harvard Smithsonian Center for Astrophysics, 60 Garden St, Cambridge, MA 02138, USA}
\altaffiltext{9}{Department of Astronomy and Institute of Theoretical Physics and Astrophysics, Xiamen University, Xiamen, Fujian 361005, China} 
\altaffiltext{10}{Department of Physics and Astronomy, University of Utah, 115 S. 1400 E., Salt Lake City, UT 84112, USA}
\altaffiltext{11}{Center for Exoplanets and Habitable Worlds, The Pennsylvania State University, 525 Davey Laboratory, University Park, PA 16802, USA}
\altaffiltext{12}{Departamento de Fisica, Facultad de Ciencias, Universidad de La Serena, Cisternas 1200, La Serena, Chile}

\begin{abstract}
We present results from the largest systematic investigation of broad absorption line (BAL) acceleration to date. We use spectra of 140 quasars from three Sloan Digital Sky Survey programs to search for global velocity offsets in BALs over timescales of $\approx$2.5--5.5 years in the quasar rest frame. We carefully select acceleration candidates by requiring monolithic velocity shifts over the entire BAL trough, avoiding BALs with velocity shifts that might be caused by profile variability. The \civ \ BALs of two quasars show velocity shifts consistent with the expected signatures of BAL acceleration, and the BAL of one quasar shows a velocity-shift signature of deceleration. 
In our two acceleration candidates, we see evidence that the magnitude of the acceleration is not constant over time; the magnitudes of the change in acceleration for both acceleration candidates are difficult to produce with a standard disk-wind model or via geometric projection effects. We measure upper limits to acceleration and deceleration for 76 additional BAL troughs and find that the majority of BALs are stable to within about 3\% of their mean velocities. The lack of widespread acceleration/deceleration could indicate that the gas producing most BALs is located at large radii from the central black hole and/or is not currently strongly interacting with ambient material within the host galaxy along our line of sight. 
\end{abstract}


\section{INTRODUCTION}
\label{sec:introduction}

High-velocity winds in quasars, observed via broad absorption line (BAL) features in their spectra, are a means to explore quasar physics and possible interactions between supermassive black holes and their host galaxies (e.g., \citealt{Dimatteo05}; \citealt{Moll07}; \citealt{King10}). BAL troughs are at least 2000 \kms \ wide (by definition; see, e.g., \citealt{Weymann91}), are seen across many different ionization species (e.g., \citealt{Turnshek84}; \citealt{Arav01}; \citealt{Gibson09}), and have been observed to be variable across a broad range of timescales, ranging from days to years in the quasar rest frame (e.g., \citealt{Lundgren07}; \citealt{Gibson08b}; \citealt{Capellupo12}; \citealt{Filizak13}; \citealt{Vivek14}; \citealt{Grier15}). A popular model suggests that the BAL features originate in a line-driven wind that is launched from the accretion disk (e.g., \citealt{Murray95}; \citealt{Proga00}; \citealt{Higginbottom14}), although there are genuine challenges to this model (e.g., \citealt{Arav13}; \citealt{Baskin14}). Different modes of BAL variability (or lack thereof), such as changes in strength, changes in profile shape, or shifts in velocity of the entire BAL, can reveal different aspects of the environment, geometry, distance, and dynamics of the wind and provide information from which to build our understanding of quasar winds and their effects on both the quasars and their host galaxies (e.g., \citealt{Barlow93}; \citealt{Capellupo12}; \citealt{Filizak13}). 

Strength and/or profile variability in BALs is widespread: 50--60\% of \civ \ and \siiv \ BALs are found to vary on timescales of a few years (e.g., \citealt{Filizak13}) and BALs of other species such as \aliii \ and \feii \ have also been seen to vary (e.g., \citealt{Vivek12b}). However, reports of monolithic velocity shifts (the implied cause of which is the acceleration of the outflow material) have appeared only a few times in the literature (\citealt{Vilkoviskij01}; \citealt{Rupke02}; \citealt{Gabel03}; \citealt{Hall07}; \citealt{Joshi14}). Studies of general BAL variability in samples of quasars by \cite{Gibson08b} and \cite{Capellupo12} find no cases of velocity shifts/acceleration, although small sample sizes limit the statistical constraints on acceleration yielded by these studies. Additionally, studies of single objects such as the investigation of NGC\,4051 by \cite{Kaspi04} often find no significant acceleration in absorption features, although in some cases they are able to place physically interesting upper limits on the magnitude of acceleration. 

An observed velocity shift of a BAL could be produced by a few different mechanisms: 1) Actual increase or decrease in the speed of the material from an intermittent outflow that is consistently moving in the same direction (e.g., \citealt{Hall07});  2) A directional shift in the outflow, resulting in a change in the observed line-of-sight velocity rather than a change in the actual speed of the outflowing material (e.g., \citealt{Hall02}; \citealt{Gabel03}; \citealt{Hall13}); or 3) Changes in velocity-dependent quantities such as absorbing gas ionization state, or column density coverage of an inhomogeneous source (due to gas transverse motion), causing the line centroid to shift. Investigating BAL velocity shifts can provide constraints on quasar outflow models. 

Deceleration of quasar winds is plausibly expected in galactic feedback models (e.g., \citealt{Silk98}; \citealt{Fabian02}; \citealt{Dimatteo05}). Interaction between outflows/winds and ambient material in the host galaxies is a key element of these models, and possible interactions between quasar winds and host galaxies have also been suggested by some observations (e.g., \citealt{Feruglio10}; \citealt{Liu13}; \citealt{Veilleux13}; \citealt{Genzel14}; \citealt{Leighly14}). As such, observations of decelerating BALs (or lack thereof) could provide evidence for (or against) these winds as agents of galactic feedback. Additionally, a lack of observed deceleration might indicate that winds are observed very close in to the central black hole; winds observed at very small radii might not yet be interacting with host material. 

Observations of acceleration and deceleration of BALs are complicated by several factors. First, acceleration signatures are expected to be small over short timescales, at least for luminous quasars; thus long time baselines are beneficial to detect velocity shifts with available instrumentation, as we expect the velocity shifts to ``build up" over time. Second, it is difficult to distinguish true acceleration from velocity-dependent profile variability that could mimic an acceleration signature. In order to identify acceleration unambiguously, at least three epochs spanning a long time baseline are required; it is plausible for a BAL to vary in such a way that it produces an observed velocity shift once, but it would be remarkable for variability to occur multiple times while preserving the shape of the BAL. Thus, if one observes a BAL that has shifted in velocity between the first and second epoch that continues to shift in velocity in a similar fashion in subsequent epochs, reasonable explanations that do not involve the acceleration of the BAL are few. 

A search for ``actual acceleration" of outflow material via velocity shifts of BALs is limited by the ability to disentangle acceleration from other forms of variability. As noted above, for example, many BALs are highly variable in profile shape, and this profile variability has the potential to render actual acceleration signatures unobservable by interfering with measurements of line shifts. A lack of detected acceleration thus does not necessarily indicate that acceleration is not occurring. For example, a lack of detected acceleration could arise if an outflow is in a stable, ``standing-flow" configuration, where the gas is continually replenished at a constant velocity (see, e.g., Figure 10 of \citealt{Arav99}). This would result in the onset velocity and profile of the BAL trough appearing constant even though the gas itself is accelerating (e.g., \citealt{Murray95}; \citealt{dekool97}; \citealt{Proga00}).

Due to small sample sizes, short time baselines, and insufficient numbers of epochs, a systematic search for BAL acceleration in quasars has not yet been performed. There have been many studies of BAL variability in samples of quasars with multiple epochs (e.g., \citealt{Lundgren07}; \citealt{Gibson08b}; \citealt{Gibson10}; \citealt{Capellupo12}; \citealt{Filizak13}); however, none of these works focused specifically on BAL acceleration --- they were primarily studies of BAL strength and profile variability, although in several cases, a lack of obvious acceleration signatures is mentioned. Table \ref{tbl:previous_studies} shows a compilation of sample size, timescales, redshifts explored, number of epochs used, and spectral resolution of these studies. 

The availability of spectra from the Sloan Digital Sky Survey (SDSS; \citealt{York00}) provides a unique opportunity to investigate systematically BAL acceleration in a large sample of quasars. Beginning in 2000, SDSS I/II obtained spectra for over 105,000 quasars with the SDSS 2.5-meter telescope (\citealt{Gunn06}; \citealt{Smee13}) at Apache Point Observatory, covering 9380 deg$^2$ of the sky (e.g., \citealt{Richards02}; \citealt{Abazajian09}; \citealt{Schneider10}). From 2008--2014, the SDSS-III Baryon Oscillation Spectroscopic Survey (BOSS;  \citealt{Eisenstein11}; \citealt{Dawson13}) targeted a sample of $\sim$210,000 quasars (\citealt{Ross12}; \citealt{Paris16}) with the goal of measuring the baryon acoustic oscillation signature in the Ly$\alpha$ forest. One of the BOSS ancillary programs was to target BAL quasars with prior SDSS observations in an effort to investigate quasar winds and BAL variability on multi-year timescales. The $\sim$2000 targets selected for this program were taken from the \cite{Gibson09b} BAL quasar catalog. These targets were chosen to be optically bright ($i < 19.3$) and to have relatively high signal-to-noise ratio spectra with prominent BAL features, and they have been the subject of a number of previous investigations of BAL variability and quasar winds (e.g., \citealt{Filizak12, Filizak13, Filizak14}).  

Autumn of 2014 marked the beginning of the SDSS-IV program; of relevance to this work is the Extended Baryon Oscillation Spectroscopic Survey (eBOSS; \citealt{Dawson16}). The main goal of eBOSS is to measure the expansion of the Universe using the BOSS spectrograph over a wider range of redshifts than has been done previously, covering 7500 deg$^2$ of the sky. The Time Domain Spectroscopic Survey (hereafter, TDSS) is a subprogram of eBOSS whose primary goal is to obtain classification spectra for $\sim10^5$ variables selected from Pan-STARRS 1 (PS1; \citealt{Kaiser02}) imaging, as detailed in \cite{Morganson15}. In addition, about 10\% of the TDSS fibers are devoted to repeated spectra of potentially time-variable targets which already have at least one earlier epoch of SDSS spectra; these cases are collectively termed the Few Epoch Spectroscopy (FES) targets in TDSS, and include both selected quasars and variable stars. The BAL sample discussed above is included in this FES sample. Adding the TDSS observations of these BAL quasars provides at least three epochs, separated by rest-frame timescales ranging from months to years, over which to investigate BAL variability and BAL acceleration; the availability of a third epoch is key in identifying velocity shifts of BALs that are due to actual acceleration rather than due to coincidental velocity-dependent variability in a single pair of epochs. 

In this work we carry out a systematic investigation of BAL acceleration/deceleration in the largest multi-epoch sample of BAL quasars to date using data from the SDSS I/II, BOSS, and eBOSS programs. To search for acceleration, we employ cross-correlation techniques commonly used with light curves of active galactic nuclei (AGNs) in reverberation mapping studies (e.g., \citealt{Peterson04}); this allows us to increase our effective velocity resolution and characterize well our measurement uncertainties. In Section~\ref{sec:data}, we present the sample of quasars, the data used in our investigation, and its preparation for analysis. We discuss our search for acceleration candidates and our derivations of upper limits on BAL acceleration and deceleration in Section~\ref{sec:accelsearch}.  A summary of our results and a discussion of the implications and physical constraints from our investigation is given in Section~\ref{sec:discussion}. We assume a cosmology with $H_{0}~=~70$~km~s$^{-1}$~Mpc$^{-1}$, $\Omega_{\rm M}~=~0.3$, and $\Omega_{\Lambda}~=~0.7$ (following \citealt{Schneider10}) throughout this work. 

\section{OBSERVATIONS AND DATA PREPARATION}
\label{sec:data} 

\subsection{Sample Selection} 
%Updated September 1
We began with the 2005 targets from the BAL catalog of \cite{Gibson09b}, which were observed by SDSS and targeted for additional observations with BOSS and TDSS. We then searched for BOSS and TDSS observations of these targets as of 2015 June 30, identifying 172 targets that were observed by all three surveys. Some of these targets had multiple observations with SDSS, BOSS, and/or TDSS; for targets with more than 3 observations, we chose the 3 observations that were spread the farthest apart in time: one from SDSS, one from BOSS, and one from TDSS. We retain additional epochs to be examined for objects of interest (although all of our acceleration candidates in Section \ref{sec:civaccel} have only 3 epochs).

\begin{figure}
\begin{center}
\includegraphics[scale = 0.47, angle = 0, trim = 0 0 0 0, clip]{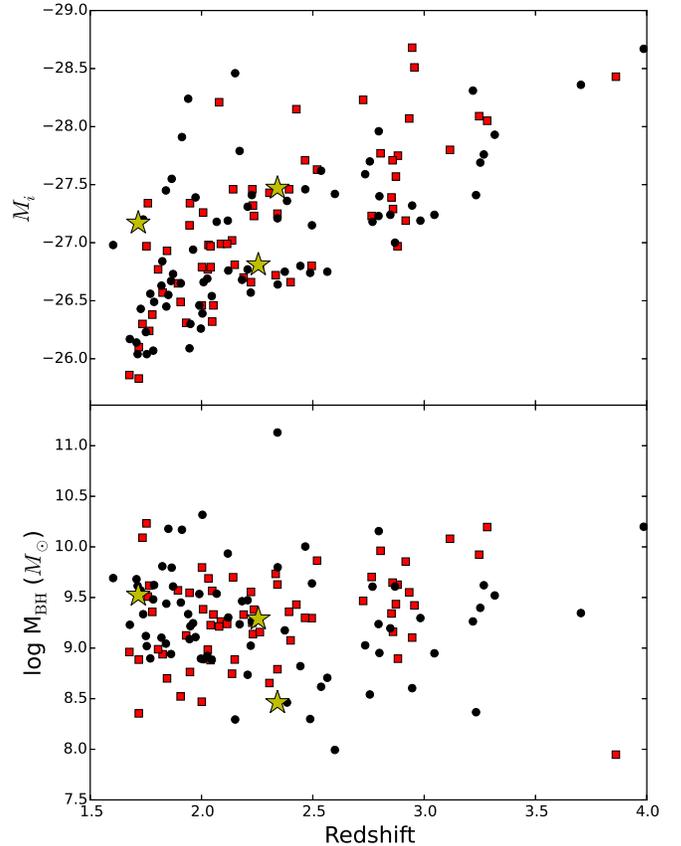} 
\caption{The distributions of $M_i$, $M_{\rm BH}$, and $z$ for our sample of BAL quasars. Yellow stars show our three acceleration candidates (see Section \ref{sec:civaccel}), red squares show objects for which we were able to place upper limits on the acceleration (See Section \ref{sec:upperlimits}), and black circles show objects for which we were unable to obtain any constraints on acceleration (Section \ref{sec:upperlimits}).}
\label{fig:mbhz}
\end{center}
\end{figure}

We restricted the redshift range of our sample to 1.5~$<~z~<~$5 to ensure good coverage of the \civ \ region, removing 11 targets from the sample. In addition, two TDSS observations appear to have ``dropped fibers" (the flux in the file was zero across the entire spectrum), leaving 159 targets. In addition, we required that all three spectra in a series have a signal-to-noise ratio (SNR$_{1700}$) equal to or greater than 6. We use the definition of SNR$_{1700}$ from \cite{Gibson09b}: SNR$_{1700}$ is taken to be the median of the flux divided by the noise per pixel over the rest-frame region from 1650$-$1750 \AA \ (at these wavelengths, one pixel spans $\sim$0.38 \AA \ in the rest frame). As this region is typically free of features and still close in wavelength to the \civ \ emission line, it provides a reasonable S/N measurement in the region of primary interest. The original BAL quasar sample from \cite{Gibson09b} was defined to include only targets with SDSS observations that had SNR$_{1700} > 6$. However, in a few cases, the BOSS and/or TDSS observations were not as high-quality, although in most cases the BOSS and TDSS observations had much higher S/N than the SDSS observations. Our SNR$_{1700}$ cut removed 19 objects from our sample, leaving 140 targets. Details on this sample are given in Table \ref{tbl:sample_info}. Most of the quasars are radio-quiet: Only 16 sources have radio flux measurements from FIRST (\citealt{White97}; compiled by \citealt{Shen11}), and the remaining 124 are undetected in the radio. Figure \ref{fig:mbhz} shows the distributions of estimated virial black hole mass ($M_{\rm BH}$; \citealt{Shen11}), absolute $i$-magnitude (\citealt{Schneider10}), and redshift (\citealt{Hewett10}) for our sample of 140 quasars. Our BAL-quasar sample spans about an order of magnitude in luminosity and two orders of magnitude in $M_{\rm BH}$. The luminosities and redshifts of our BAL quasars are representative of those of the typical BAL quasars that have been studied for decades. 

\begin{figure}
\begin{center}
\includegraphics[scale = 0.45, angle = 0, trim = 0 0 0 0, clip]{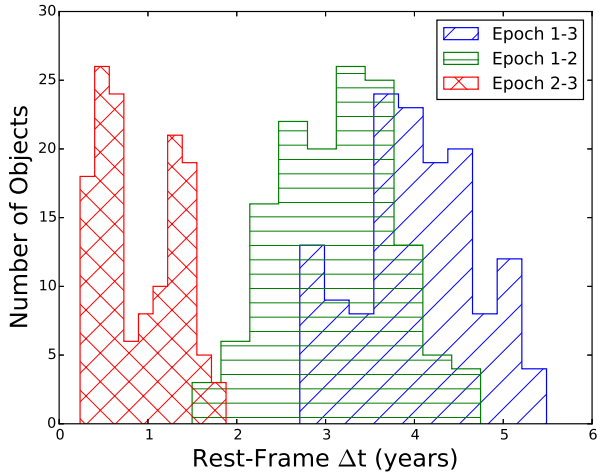} 
\caption{The distribution of rest-frame time between SDSS and TDSS (Epochs 1 and 3; blue diagonal stripes),  between SDSS and BOSS (Epochs 1 and 2; green horizontal stripes), and between BOSS and TDSS (Epochs 2 and 3; red cross-hatched) observations for the 140 targets in our sample. The BOSS and TDSS observations are generally much closer together in time than the SDSS and BOSS observations. The apparent two-peaked distribution between the BOSS and TDSS epochs is due to the scheduling patterns of the surveys on the sky, since TDSS is still ongoing.}
\label{fig:timescales}
\end{center}
\end{figure}

Figure \ref{fig:timescales} shows the distribution of the maximum $\Delta t$ (rest-frame time) spanned for each target. Largely because TDSS is still in its early stage of observations, the time between the SDSS and BOSS observations was in general significantly longer than the time between the BOSS and TDSS observations; the median timescale between the SDSS and BOSS observations is 3.14 years, while the median timescale between the BOSS and TDSS observations is 0.74 years. The median rest-frame timescale between the longest-timescale observations (i.e. between the SDSS and TDSS) is 4.05 years. Thus the time range explored by our sample (shown in Table \ref{tbl:previous_studies}) is fairly similar to the prior BAL variability studies, which explore rest-frame timescales from several days to 8.2 years.  However, our sample size is larger by almost a factor of six and our work includes a minimum of 3 epochs per quasar. 

\subsection{Data Preparation and Continuum Fitting} 
\label{sec:confits} 
Before analysis, the spectra from each survey were visually inspected. Where possible, the BOSS spectra were corrected for calibration errors that result from atmospheric differential refraction and fiber offsets during observations --- this improvement was accomplished using corrections introduced by \cite{Margala15}. We then masked all pixels from the spectra that are flagged by the SDSS, BOSS, and TDSS data-reduction pipelines as bad, usually due to sky contamination --- this action was done using the ``BRIGHTSKY" mask column of the spectral files. The processed spectra were then corrected for Galactic extinction using an $R_{\rm V}$~=~3.1 Milky Way extinction model (\citealt{Cardelli89}) and $A_{\rm V}$ values from \cite{Schlegel98}.  Before beginning the analysis, we converted the observed wavelengths of the spectra to the rest frame using redshifts from \cite{Hewett10}. 
These redshift measurements are made via cross correlation of the \mgii,  \ciii \ (for 1.5 $< z < $ 2.1), and/or  \civ \ (for 2.1 $< z < $ 4.5) emission lines. Because these redshifts rely on emission lines that often have BALs superimposed on them (particularly, in our case, the \civ \ emission line), redshifts for BAL quasars are often less reliable than for non-BAL quasars. However, all of the redshifts measured by \cite{Hewett10} for our sample were deemed reliable fits, and thus we adopt these for our work. 

Following, e.g., \cite{Gibson09b} and \cite{Filizak12}, we fit a reddened power-law model to the continuum emission using the SMC-like reddening model from \cite{Pei92} and a nonlinear least-squares fitting algorithm. We fit only regions of the spectrum that were largely devoid of strong emission and absorption features, as defined by \cite{Gibson09b}: $1250~-~1350$~\AA, $1700~-~1800$~\AA, $1950~-~2200$~\AA, $2650~-~2710$~\AA, $2950~-~3700$~\AA, $3950~-~4050$~\AA, and $4140~-~4270$~\AA. We made one update to these relatively line-free (RLF) regions -- we restricted the first RLF region to $1280-1350$~\AA, as many of our targets showed emission features toward the blue end of this first RLF region. These were used as our default RLF fitting regions; however, for the majority of our targets with $z < 1.85$, we also used an additional region from 1425--1450 \AA, which was mostly devoid of features in these targets, to help constrain the blue end of the continuum. We weighted each pixel such that each individual region contributed equally in the continuum fit to ensure that none of the RLF regions holds more weight in the fit simply because it contains more pixels. For 15 of our targets, the default RLF regions resulted in poor continuum fits due to the presence of particularly strong emission or absorption features within these windows; for these targets, the RLF regions were adjusted manually to exclude regions with strong features. This action usually resulted in the restriction or exclusion of one or two of the bluest RLF regions, which fall in the Lyman-$\alpha$ and \civ \ regions --- these regions were heavily contaminated by absorption and/or emission in these objects and sometimes showed little-to-no visible continuum to fit.   

To characterize the uncertainties in the continuum fits, we used ``flux randomization" Monte Carlo iterations (e.g., \citealt{Peterson98}) to alter the flux in each pixel of the spectrum by a random Gaussian deviate based on the spectral uncertainty. The continuum was then fit to the new spectrum and the process repeated 100 times. We then adopt the standard deviation of the 100 trials as the uncertainties of the continuum fit. 

After determining a continuum, we fit a line profile to the continuum-subtracted \civ \ emission line in each spectrum to remove the emission-line signal from our spectra --- this plus the power-law continuum becomes the overall normalizing continuum for each spectrum. The \civ \ emission-line region can be complex; it is often blended with several other emission features, and frequently has many strong absorption features superimposed. Thus, fitting an accurate profile to the \civ \ emission feature is nontrivial. 
However, we are not using the emission-line profiles to obtain any physical information, so complexities such as isolating the \civ \ line from its blended components are unnecessary for our purposes; we need only to fit the combined line profile. 
To remove the emission feature, we tried three different types of line profiles for each spectrum: 1) A Voigt profile (following \citealt{Gibson09b}) which yields a symmetric emission line, 2) A double-Gaussian profile (e.g., \citealt{Park13}), and 3) a Gauss-Hermite profile (see Appendix A of \citealt{vanderMarel93}). 
For this last option, we began with a 6th-order Gauss-Hermite profile (e.g., \citealt{Assef11}; \citealt{Denney12}), but in nearly all cases this resulted in unreasonable fits due to absorption features; however, a 4th-order Gauss-Hermite profile generally yielded satisfactory fits, so we adopt this instead.  

For all profile fits, we used an iterative fitting technique (e.g., \citealt{Gibson09b}) to ignore wavelength bins that differed by more than 3$\sigma$ from the fit, which generally helped exclude the absorption features that overlap with the emission line. We used an automatic fitting algorithm to fit the profiles, allowing the center of the profile to vary; however, in many cases, there was such strong overlapping absorption on the \civ \ emission line that an automated fit did not yield a reasonable result. 
To remedy this situation, we manually excluded regions from the fit that suffered from heavy absorption when necessary. We visually inspected each emission-line fit and chose the best line profile (Voigt, double-Gaussian, or Gauss-Hermite) for each target, requiring the same type of profile for all three epochs of each target. In all cases, it was straightforward to visually identify the best fit by examining the overall fit and residuals of all three epochs. 
In about half of the targets, a double-Gaussian yielded the best fit to the \civ \ emission line, while about a third of the targets were best fit with a 4th order Gauss-Hermite profile and the rest were best fit with a Voigt profile. In three cases, there was no emission line present due to overwhelmingly strong absorption features. The best-fit profiles are listed by object in Table \ref{tbl:sample_info}.

The spectra were then divided by the combined continuum and \civ \ model to obtain a set of ``normalized spectra" to examine. The uncertainties from both the continuum + emission-line fit and the spectrum itself were propagated to determine the final uncertainty on the normalized spectrum. The continuum + emission-line normalization accounts for intrinsic variations in the continuum and also allows a search for possible wavelength shifts of low-velocity BALs that are superposed on the \civ \ line without emission-line interference (see Section \ref{sec:ccfsec} below).  

\section{BAL Acceleration Investigation} 
\label{sec:accelsearch} 
\subsection{BAL Trough Complex Identification} 
\label{sec:troughs}
For a quasar to be included in our study, we required that it have at least one absorption trough in the \civ \ region (spanning outflow velocities between 0 and 30,000 \kms) that is wider than 2000 \kms \ in at least two of the three epochs (therefore fulfilling the formal criteria of \citealt{Weymann91} to be considered a BAL in these epochs) and does not disappear entirely in the third. To search for BAL troughs, we first smoothed the spectra by 3 pixels (which reduces the resolution to $\Delta v~\sim~$207~\kms). We then identified each BAL trough in each spectrum, defined by a region for which the quasar flux is continuously under or equal to 90\% of the continuum flux. The minimum and maximum velocities spanned by the BAL, \vmin and $v_{\rm max}$, were measured for each trough as the outflow velocity where the normalized flux reaches 90\% of the continuum flux on either side of the trough.  In 19 of our targets, there was no BAL present that met our requirement for consideration (the BAL either disappeared or weakened significantly such that it is no longer formally considered a BAL in at least two of the epochs); while these 19 targets are interesting for other reasons (see, e.g., \citealt{Filizak12}), they are not appropriate for a study of BAL acceleration, so we defer their discussion to future work. We removed these targets from our sample, leaving us with 121 quasars. 

\begin{figure}
\begin{center}
\includegraphics[scale = 0.38, angle = -90, trim = 0 0 0 0, clip]{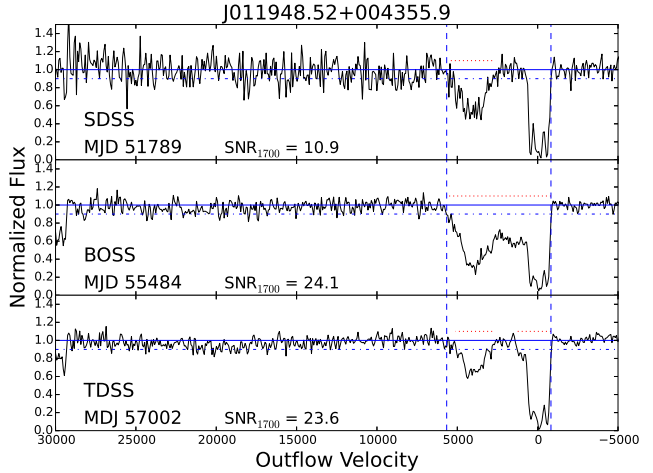} 
\caption{A set of three spectra for one of our targets, shown to demonstrate our trough-identification algorithm. The normalized spectra of all three epochs are shown as black solid lines. The solid blue line represents a normalized flux of 1.0, the dashed-dotted blue line is a flux level of 0.9, and the red dotted lines denote each contiguous region of the normalized spectrum where the normalized flux remains below 0.9 for a duration that is wider than 2000 \kms. The dashed blue lines display the final adopted \vmin and \vmax for the BAL-trough complex. To demonstrate better the data quality, the spectra shown here are not smoothed; however, as noted, when identifying the BAL trough complexes, we smoothed the spectra by three pixels. }
\label{fig:complexes}
\end{center}
\end{figure}

Because we are investigating multiple epochs of data for each target, BAL identification can be complex; BAL troughs do not necessarily remain isolated throughout time. In some cases, a single BAL trough in one epoch may appear separated into multiple distinct troughs in another epoch; we consider these troughs a part of a larger BAL ``complex". Following \cite{Filizak13}, we identify BAL complexes and treat each complex as an individual trough entity in our subsequent analysis. To identify BAL complexes, we started with our initial set of BAL troughs (defined as those spanning at least 2000 \kms \ in at least two epochs, as discussed above) and their \vmin and \vmax values at each epoch. We then identified complexes as structures where multiple troughs in some epochs correspond to a single, wider trough in a different epoch --- this phenomenon occurred in $\sim$10\% of our targets. We defined \vmin and \vmax of the complex at the most extreme limits over the three epochs so that the velocity region taken into account includes the entire system in all three epochs.  A sample set of observations for one of our targets is shown in Figure \ref{fig:complexes} to demonstrate our trough-complex identification algorithm.
Most of our targets show only a single \civ \ BAL trough or trough complex, although some have multiple complexes in the \civ \ region; we identified 151 different \civ \ BAL trough complexes in our final sample of 121 targets. Our search began by covering the range 0~$<~v~<~$30000 \kms; however, in some cases, the BALs extended slightly below 0 \kms \ (i.e., to gas that is redshifted relative to the quasar emission lines), so we allowed our BAL measurements to extend a few hundred \kms \ below zero when necessary to avoid truncating them. We do not explore BALs at velocities greater than 30,000 \kms \ in this work. 

\subsection{Cross-Correlation Function Analysis}
\label{sec:ccfsec} 

A main goal of our study is to improve the methodology used to search for acceleration; as the first large systematic study of BAL acceleration, we aim to provide better precision and accuracy than previous works that have discussed BAL acceleration. To our knowledge, the only other studies to discuss acceleration in samples of multiple quasars are \cite{Gibson08b, Gibson10}, and \cite{Capellupo12}. \cite{Gibson08b} assumed an upper limit of 1 \AA \ on the velocity shifts for each source and obtained upper limits on acceleration by dividing by the relevant timescale between epochs for each source. \cite{Gibson10} discuss manual shifting and visual inspection to identify possible acceleration in two targets, and \cite{Capellupo12} do not discuss the method by which they determined there was no acceleration detected in their sample. 

In our study, we strive to obtain tighter constraints with more rigorous quantification of uncertainties. To do so, we adopt a standard cross-correlation technique often employed to measure time delays between light curves in reverberation mapping (e.g., \citealt{Peterson04}) as well as in searches for exoplanets using the radial-velocity method (e.g., \citealt{Latham89}). We use the cross-correlation function (CCF) method originally introduced by \cite{Gaskell86} and \cite{Gaskell87}; these techniques have been tested and improved over the years. CCF methodology allows an improvement in our sensitivity to small velocity shifts compared to methods using integer-pixel shifts, as it is possible to use the centroid of the CCF to measure shifts of less than a pixel in some cases. The magnitude of the uncertainties in the velocity shifts measured via CCF depend on the specific absorption feature in question: Trough depth, velocity width, and steepness of the profile determine the size of uncertainties, along with data-quality issues such as S/N and spectral resolution (see \citealt{Beatty15} for a thorough discussion of the effects of trough shape/width and data quality on uncertainties). The CCF methodology as discussed below allows placement of robust constraints on the uncertainties in our measurements, allowing measurement of robust upper limits in cases where no significant acceleration is detected. 

We first converted all spectra into velocity-space instead of wavelength-space by calculating the outflow velocity at each pixel (using the rest-frame wavelength of \civ \ as the zeropoint) and cropped the spectra to include only the region spanning the specific BAL trough complex in question plus 2000 \kms \ of padding on each end. In cases where other absorption features or bad pixels were present within this 2000 \kms \ padding window, we manually altered the padding to contain as few extraneous features as possible. The CCF works as follows: We cross-correlate the two spectra and calculate the value of Pearson's cross-correlation coefficient $r$. The first epoch is then shifted by a velocity unit (in this case, we shift by one pixel, or 69 \kms), and the cross-correlation coefficient is recalculated. We performed this exercise across velocity shifts ranging from $-$2000 \kms \ to $+$2000 \kms \ to build the CCF, which consists of the value of $r$ for each velocity shift explored. We measure the velocity shift that results in the highest, or peak, correlation coefficient, as well as the centroid of the CCF about the peak, calculated using points surrounding the peak with values greater than 0.8$r_{\rm peak}$. 

\begin{figure}
\begin{center}
\includegraphics[scale = 0.46, angle = -90, trim = 0 0 0 0, clip]{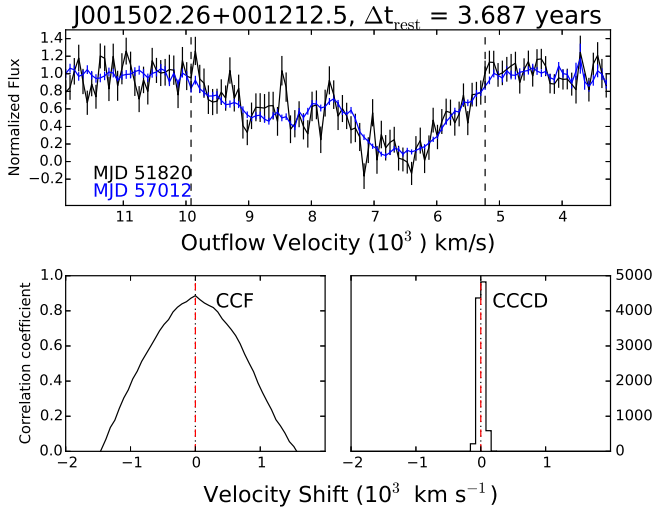} 
\caption{A sample cross-correlation function for two spectra in our study. The top panel shows the two spectra in black (first epoch; in this case, SDSS) and blue (second epoch; in this case, TDSS). The vertical dashed lines denote \vmin and \vmax of the BAL. The bottom panels present the results from the CCF analysis: The bottom-left panel is the CCF for these two spectra and the bottom-right panel is the cross-correlation centroid distribution (CCCD; as measured from the Monte Carlo simulations). The $y$-axis on the bottom-right panel displays the number of trials from the Monte Carlo iterations. The vertical red dashed lines in the bottom panels indicate the centroid shift measured from the actual data rather than the median from the Monte Carlo iterations. To help guide the eye, a shift of 0 \kms \ is represented by the black dotted line in the bottom panels. In this particular example, we do not measure a significant velocity shift.}
\label{fig:ccf}
\end{center}
\end{figure}

To obtain the best measurement of the velocity shift and its uncertainties, we use Monte Carlo simulations that employ the flux randomization method of \cite{Peterson98}, refined by \cite{Peterson04}. In each simulation, the fluxes of both spectra are altered by a random Gaussian deviate associated with each pixel's spectral uncertainties. The CCF is then recalculated for each realization of the spectra and the centroid and peak of the CCF are calculated; this is repeated 10,000 times. We adopt the median of the distribution of CCF centroid measurements from the 10,000 iterations as the best velocity shift, the velocity shifts corresponding to the 68.3\%  percentile of the cross correlation centroid distribution (CCCD) as our formal 1$\sigma$ uncertainties, and the velocity shifts corresponding to the 99.73\% percentile of the CCCD as our formal 3$\sigma$ uncertainties. An example of a CCF, CCCD, and cross-correlation peak distribution (CCPD) are shown in Figure \ref{fig:ccf}. 

When using the CCF to measure shifts in spectra, the difference between the centroid of the CCF in the actual data and the median centroid from the Monte Carlo iterations is negligible because of the regular sampling of the spectra in wavelength space; however, this method was originally developed for use with light curves (e.g., reverberation mapping; see \citealt{Peterson04}) where the sampling is often irregular and sparse with large uncertainties, and in such cases the median of the distribution of lags often yields a better characterization of the shift. We here adopt the median, and note that the results using the centroid from the actual data do not change our results (see Figure \ref{fig:ccf}). Typical 1$\sigma$ uncertainties on velocity shifts for our sample are on the order of about half of a pixel in size (we find a median uncertainty of $\sim$33~\kms), but range from uncertainties as low as one tenth of a pixel to as high as two pixels in extreme cases. Typical 3$\sigma$ velocity-shift uncertainties are on the order of 1--2 pixels in size, again with a wide range based on the quality of the spectrum and the details of the trough shape. We compared our measured uncertainties to those predicted by \cite{Beatty15} when measuring velocity shifts in stellar spectra and find that our measured uncertainties are consistent with predictions to within a factor of a few (some deviation is expected, as the predictions are made for more ideal spectra than our data, and BAL troughs are often quite a bit more complex in profile than the stellar absorption features used in radial-velocity surveys). 

After calculating the CCF between all pairs of spectra for each trough in each target, we then searched for pairs of spectra that had a measured centroid shift that was at least greater than its 3$\sigma$ uncertainty (the shift was measured at $\geq$ 3$\sigma$ significance). Targets that do not show a measured centroid shift at $\geq3 \sigma$ significance are considered non-detections and provide upper limits; the upper limit measurements are discussed in Section \ref{sec:upperlimits}. All measured velocity shifts and their uncertainties were converted to units of acceleration/deceleration by dividing the velocity shift by the rest-frame time between the two epochs. We note, however, that this procedure assumes that the acceleration is constant over the measured period, which is not necessarily the case (See Section \ref{sec:civaccel} for further discussion). 

Hereafter, when discussing general strategies for our search, we refer to both positive acceleration (shifts toward higher outflow velocities) and negative acceleration (shifts toward smaller outflow velocities) as simply ``acceleration". In cases where we need to differentiate between acceleration in the negative direction and in the positive direction, we refer to positive acceleration (shifts toward higher outflow velocities) as ``acceleration" and negative velocity shifts (shifts toward smaller velocities) as ``deceleration". 

\subsection{BAL Profile Variability Tests} 
\label{sec:shapetest} 
In order to measure possible acceleration  of BAL profiles, we must be able to distinguish a true acceleration signature from other types of BAL variability. In cases where the overall strength and/or shape of the BAL changes dramatically over the course of our observations, measuring acceleration becomes more complicated. To obtain a sample of robust acceleration candidates, we first isolate those BAL troughs that do not change over the campaign from those BALs whose profiles change significantly in shape and/or strength. To make this selection, we applied the measured velocity shift to the first (earlier) spectrum in each pair of epochs to align them with the BAL in the subsequent epoch. We then searched for shape and/or strength changes using a $\chi^2$ test: If the BAL shifted only in velocity, applying the shift measured by the CCF should result in a good match (and thus a low $\chi^2$) between the BAL profiles in the two epochs; if the BAL profile itself varied in shape or strength, even applying a measured ``shift" will not result in a good match. 

We calculated the $\chi^2$ statistic between the two spectra before and after the velocity shift was applied, using \vmin and \vmax of the BAL trough to define the range for this calculation. We first linearly interpolated the shifted spectrum so that the velocities/wavelengths of the shifted spectrum matched those of the spectrum to which it was being compared. We then calculated the reduced $\chi^2$ and corresponding $p$-value for each absorption profile both before and after shifting by the velocity shift measured by the CCF. To obtain the average uncertainties for use in the $\chi^2$ calculation, we added together the uncertainties of the two spectra in quadrature (see \citealt{Bevington03}). To be considered an acceleration candidate, two requirements were imposed: 1) The $p$-value comparing the {\it unshifted} spectrum with the subsequent epoch was $<$~0.1 (we require a significant difference between the two spectra initially), and 2) When calculating $\chi^2$ between the {\it velocity-shifted} first epoch and the subsequent epoch, the $p$-value is $\geq$  0.1 (i.e., the probability of obtaining the measured $\chi^2$ in two troughs that are actually different is less than 10\%). 

The cases where the CCF identifies a significant shift but the first requirement ($p <$ 0.1 before applying the shift) is not met are few, and thus it is used primarily as a sanity check. In the four cases where the CCF does measure a 3$\sigma$ shift but the unshifted $p$-value indicates a good fit, the issue arises because the wavelength range used in the CCF is slightly different from that used in the $\chi^2$ calculation. As noted above, when calculating the CCF, we used the entire trough plus 2000 \kms \ on either end to allow for the wavelength shifts. However, when calculating the $\chi^2$, we consider only the spectrum between $v_{\rm min}$ and $v_{\rm max}$. Cases where additional spectral features (either absorption or bad pixels) are found within the padding region on either side of the BAL can cause the CCF to identify a shift that is caused by the extraneous spectral features rather than by a shift of the BAL itself. We attempted to remedy this by manually adjusting the padding regions in cases where other features were visible within this window, but we were unable to exclude all the features in some cases, so our first $p$-value requirement helps correct for this. The second $\chi^2$ requirement ($p \geq $ 0.1 after applying the shift) is necessary to avoid variability in BAL shape and/or strength, where a significant velocity shift reported by the CCF may be solely due to velocity-dependent variability within the BAL. Changes in trough shape are easily identifiable by visual inspection or applying our $\chi^2$ test, but the CCF alone is not sufficient to distinguish between monotonic velocity shifts and velocity-dependent variability. 

If these criteria are met for the pair of spectra spanning the longest time baseline (the SDSS and TDSS epochs), we consider that object as an ``acceleration candidate" and discuss it in detail below. We note that by applying a requirement that the BAL line profile is not variable, we are possibly excluding BALs where genuine acceleration has occurred; in some cases, the BAL could have both accelerated and varied in strength and/or shape, and thus imposing this requirement will cause us to miss these cases. However, due to the difficulties in disentangling these two effects with only three epochs, we primarily focus in this study on BALs whose strength or profile shape did not change significantly over the period of observations --- we are searching for unambiguous, ``clean" examples of acceleration. For targets with BALs that do change in shape and/or strength or those for which we did not measure a significant velocity shift, we set upper limits on BAL acceleration (see Section \ref{sec:upperlimits}).  

\subsection{Acceleration Candidates}  
\subsubsection{\civ \ BALs} 
\label{sec:civaccel}

\begin{figure*}
\begin{center}
\includegraphics[scale = 0.46, angle = -90, trim = 0 0 0 0, clip]{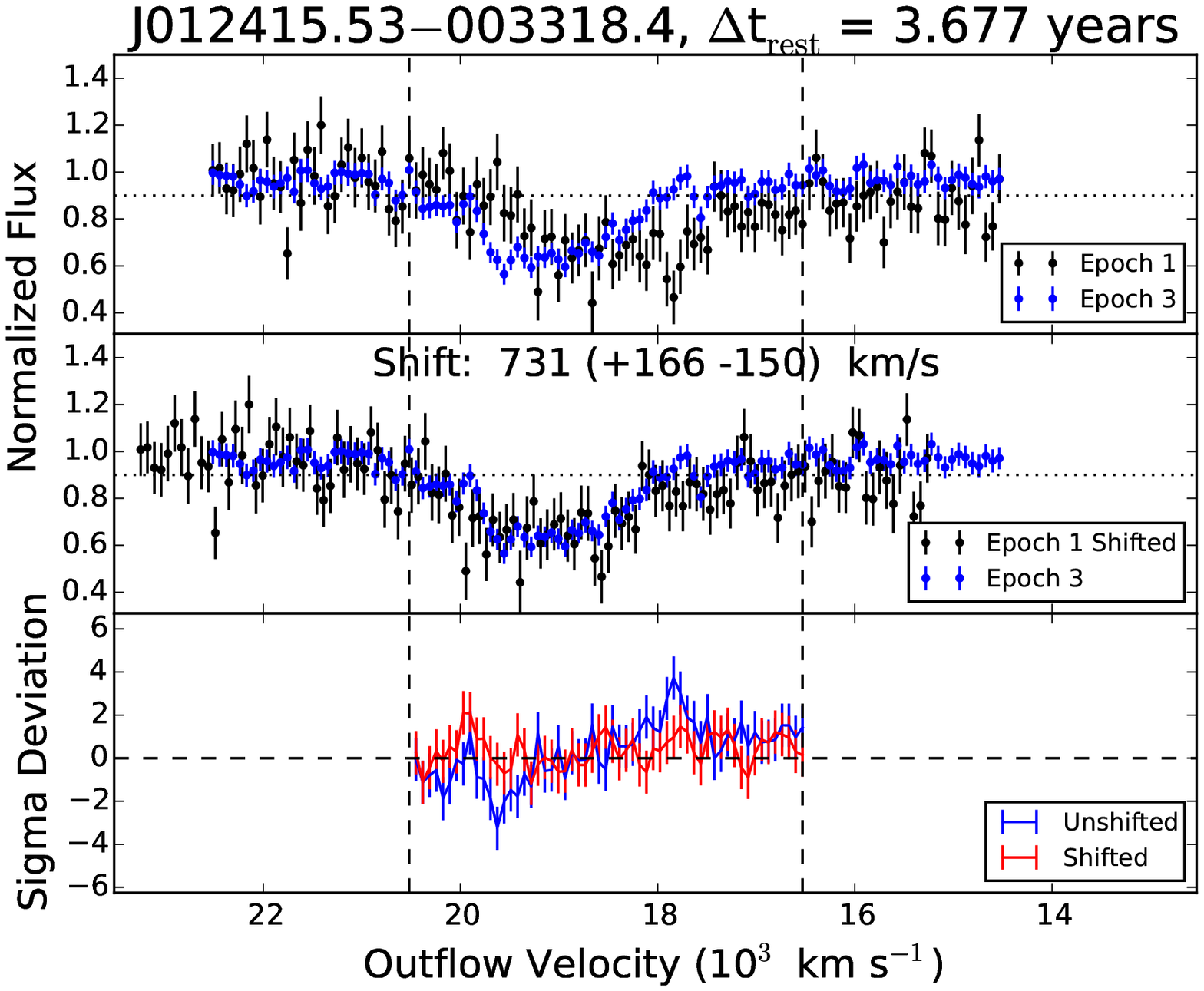} 
\includegraphics[scale = 0.46, angle = -90, trim = 0 0 0 0, clip]{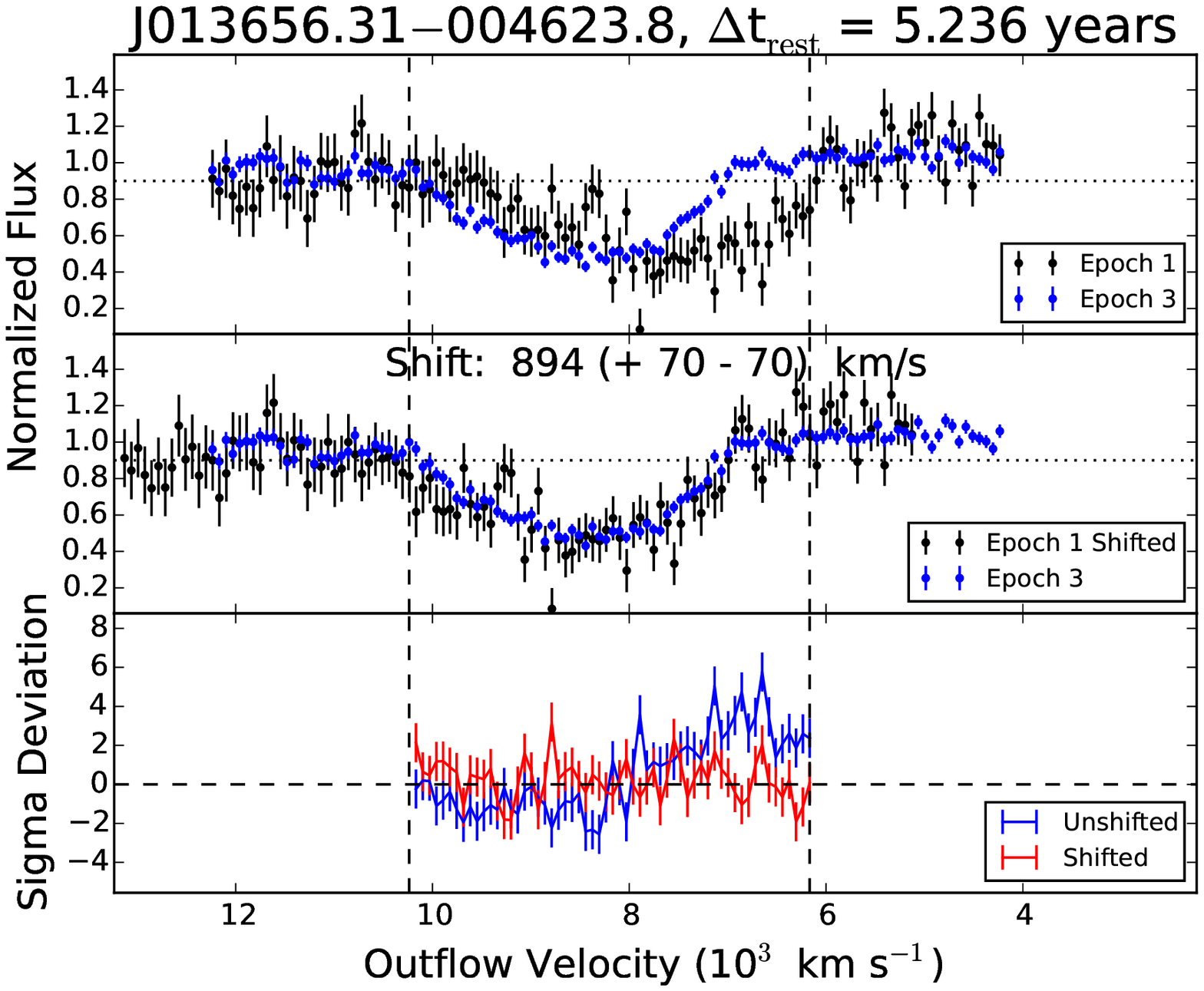} 
\includegraphics[scale = 0.46, angle = -90, trim = 0 0 0 0, clip]{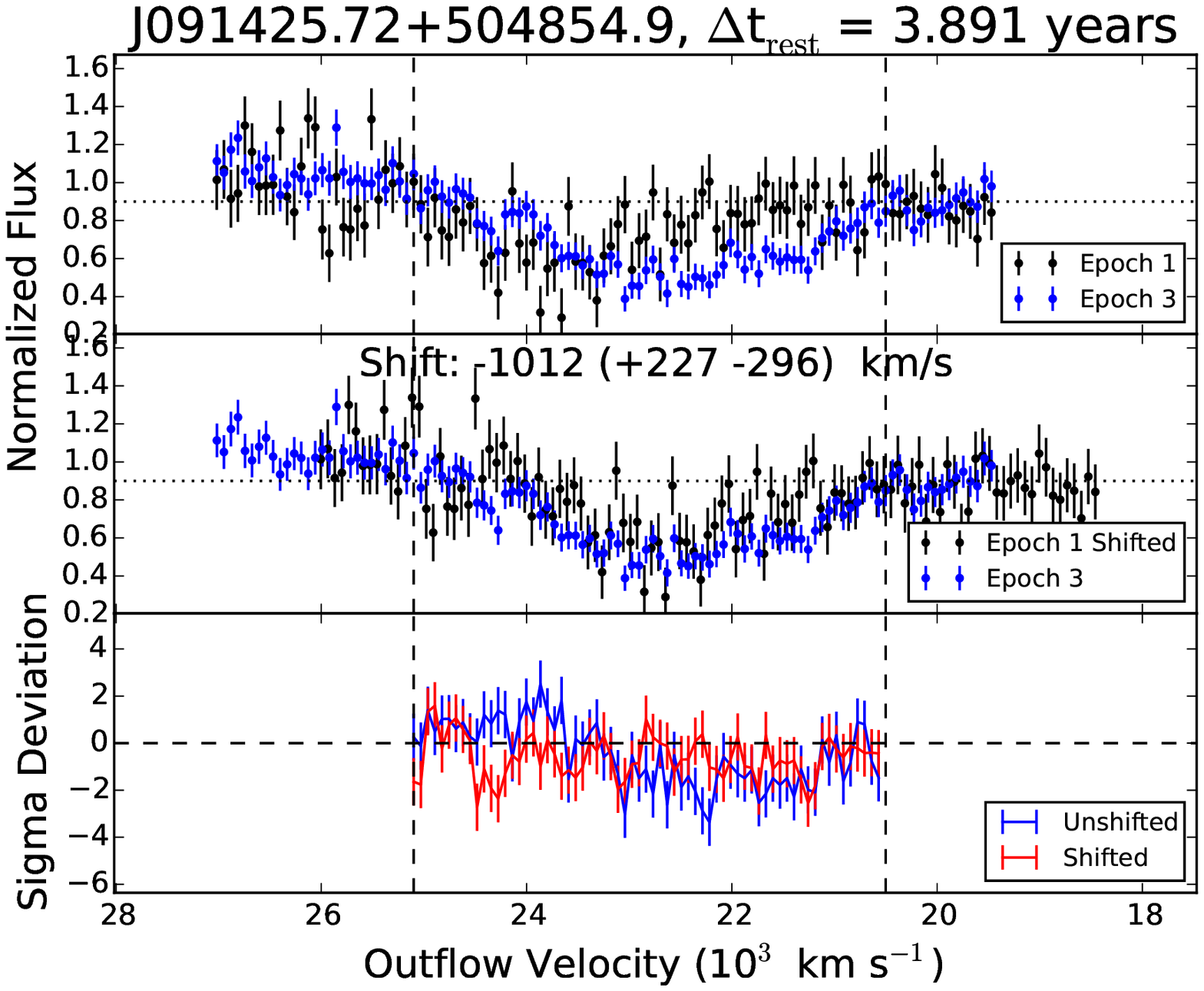} 
\caption{The \civ \ BALs of our acceleration candidates. In each case, the first (SDSS) epoch is shown in black, and the additional epoch (TDSS) is shown in blue. The middle sub-panel of each panel presents the same two epochs with the SDSS epoch shifted by the measured velocity shift from the CCF analysis. The bottom sub-panel displays the sigma deviation between the spectra before and after the shift in red and blue, respectively. }
\label{fig:all_candidates}
\end{center}
\end{figure*}

\begin{figure*}
\begin{center}
\includegraphics[scale = 0.44, angle = -90, trim = 0 0 0 0, clip]{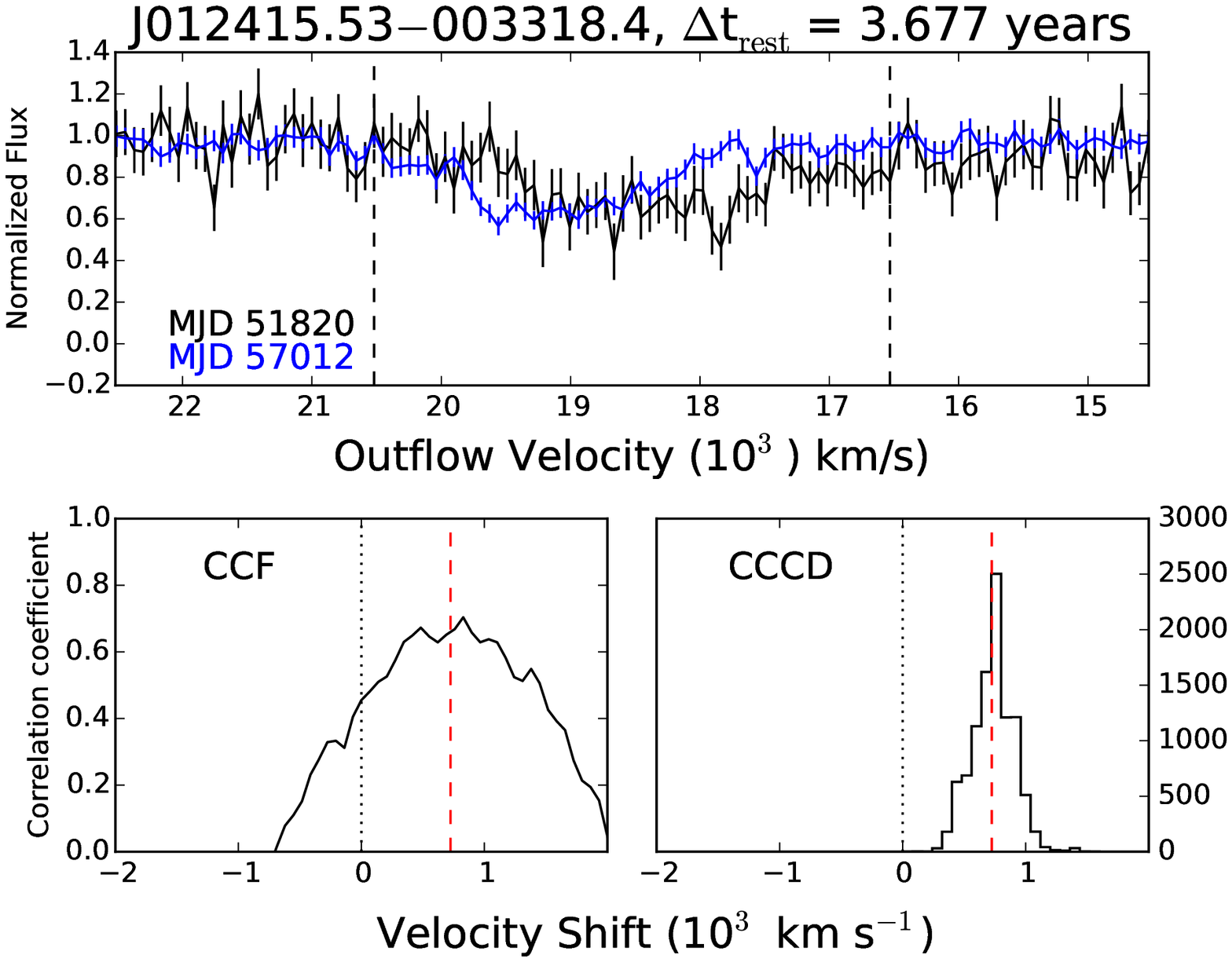} 
\includegraphics[scale = 0.44, angle = -90, trim = 0 0 0 0, clip]{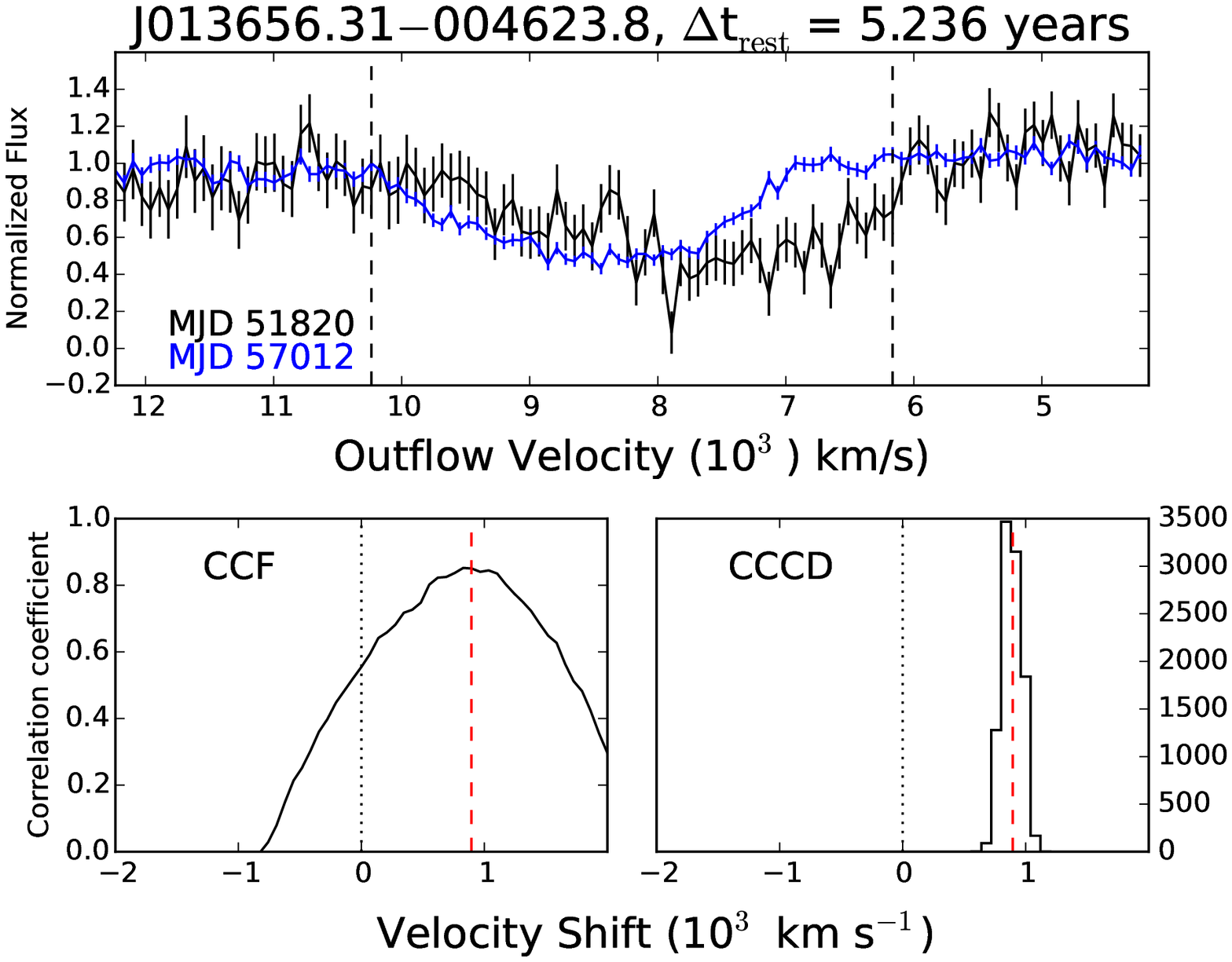} 
\includegraphics[scale = 0.44, angle = -90, trim = 0 0 0 0, clip]{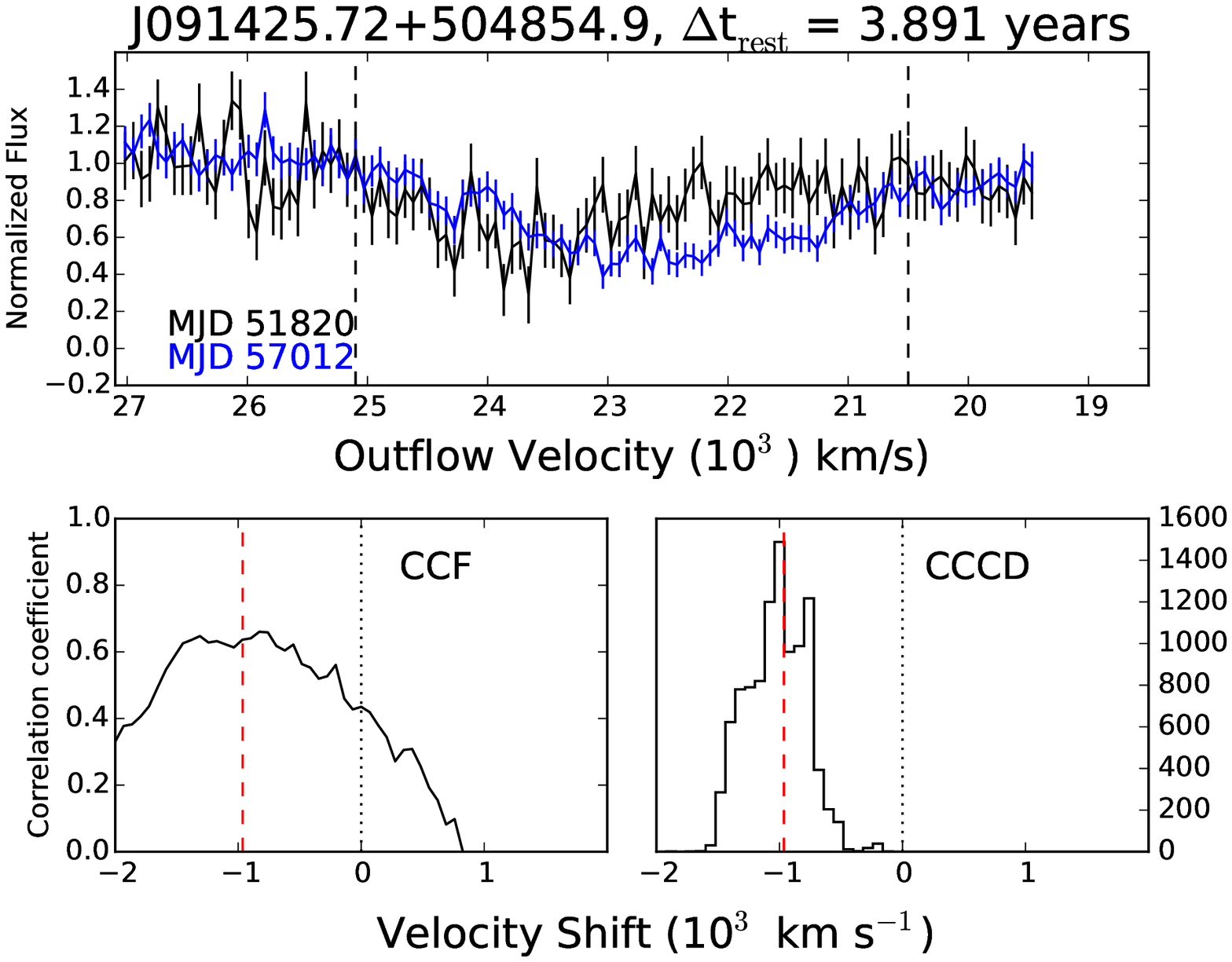} 
\caption{Cross-correlation functions for the three acceleration candidates. The top subpanel for each candidate shows the two spectra in black (first epoch; in this case, SDSS) and blue (second epoch; in this case, TDSS). The vertical dashed black lines denote \vmin and \vmax adopted for the BAL. The bottom subpanels in each panel present the results from the CCF analysis: The bottom-left subpanel is the CCF for the two spectra and the bottom-right subpanel is the cross-correlation centroid distribution (CCCD; as measured from the Monte Carlo simulations). The $y$-axis on the bottom-right subpanel displays the number of trials from the Monte Carlo iterations. The vertical red dashed lines in the bottom subpanels indicate the centroid shift measured from the actual data rather than the median from the Monte Carlo iterations. To help guide the eye, a shift of 0 \kms \ is shown by the black dotted lines in the bottom subpanels.}
\label{fig:candidate_ccfs}
\end{center}
\end{figure*}

After applying these requirements, we find three candidates for BAL acceleration in our sample. The measured $v_{\rm min}$, $v_{\rm max}$, velocity shift, measured average acceleration, and $\chi^2$ test results for each pair of epochs for each candidate are listed in Table \ref{tbl:candidates}. Figure \ref{fig:all_candidates} shows \civ \ BAL spectra for the first and third (SDSS and TDSS) observations of each of these candidates and Figure \ref{fig:candidate_ccfs} shows the CCF results between these two epochs. A detailed discussion and evaluation of the individual candidates is given in the Appendix.  Two of the three targets show an increase in outflow velocity of the BAL over time (J012415.53$-$003318.4 and J013656.31$-$004623.8), with average acceleration magnitudes of 0.630$^{+0.14}_{-0.13}$~cm~s$^{-2}$ and 0.541$\pm$0.04~cm~s$^{-2}$ over the span of the three observations. In one target (J091425.72+504854.9), we see evidence for {\it deceleration} of the BAL at an average rate of $-0.83^{+0.19}_{-0.24}$~cm~s$^{-2}$. Following the discussion in the Appendix, we consider the cases of J012415.53$-$003318.4 and J013656.31$-$004623.8 to be our strongest acceleration candidates; it would be remarkable for a BAL to vary in a velocity-dependent way that results in such a stable profile shape while simultaneously appearing to change in velocity. Due to the low S/N and shallower BAL profiles, the deceleration case of J091425.72+504854.9 is slightly less certain (see Appendix for a more thorough discussion). Two of the three candidates do not have multiple BALs in the \civ \ region, while one (J091425.72+504854.9) has four separate BAL troughs in this region. In this case, we do not detect acceleration in the additional \civ \ BALs. This could indicate that the \civ \ BALs at different velocities originate from winds that are not co-located, and instead represent distinct flows/streams of gas. 

In all cases, the acceleration signature in the \civ \ BALs is less robust between the BOSS and TDSS epochs than it is between the SDSS and BOSS epochs because of the shorter timescales between the BOSS and TDSS observations (see the Appendix for details). While we do measure a velocity shift that is significant at greater than 3$\sigma$ between the BOSS and TDSS epochs for all candidates, at least some variability in strength and/or shape of the profile is detected between these epochs in all three cases. It is possible that small changes in shape occurred between the SDSS and BOSS epochs as well, but that the lower S/N of the SDSS data masks this effect. 

In our two most robust cases (J012415.53$-$003318.4 and J013656.31$-$004623.8), the magnitude of the acceleration measured between Epochs 1 and 2 is not consistent with that observed between Epochs 2 and 3; the acceleration magnitude decreases in both cases. This change in acceleration magnitude was also reported by \cite{Gabel03} in a narrower absorption system; this result suggests that whatever acceleration we are observing is not constant over time. Thus our assumption of constant acceleration in our calculations is likely false; instead, the acceleration measurements obtained over the time spanned between epochs represent the {\it average} acceleration over that period --- while our third epoch is instructive, we require additional epochs to obtain information on how the magnitude of the acceleration evolves over time (i.e., to measure the jerk). 

A possible source of systematics in our analysis is the \civ \ emission-line fit in each quasar --- because the \civ \ emission line is variable, we fit each epoch individually. As discussed in Section 
\ref{sec:confits}, the \civ \ emission-line profiles are often complex, and it can be difficult to disentangle them from absorption features as well as the continuum. Acceleration candidates in J012415.53$-$003318.4 and J091425.72+504854.9 are sufficiently detached from the \civ \ emission line that there is no overlap; hence our emission-line fits do not affect the analysis for these targets. The BAL in J013656.31$-$004623.8 is at a lower velocity and is not detached from the emission line. However, the observed acceleration in J013656.31$-$004623.8 is still detected when the spectra are normalized by the continuum only (without an emission-line fit), although the CCF and $\chi^2$ analysis is difficult to perform without a flat continuum on either side of the emission line. 

As we have set our threshold detection significance level at 3$\sigma$, which corresponds to 99.73\% significance, we expect 0.40 false positives in a sample of 151 BALs. This is significantly smaller than the observed detection rate of three; we are thus confident that our three candidates are overall unlikely to be false detections. Although three acceleration candidates is a small number for robust statistics, we also searched for preliminary signs of trends in the properties of both the quasars hosting acceleration candidates and the properties of the BALs. Figure~\ref{fig:mbhz} shows that the three BAL-acceleration candidates are present in quasars with fairly typical properties (they are not outliers in luminosity or redshift compared to the rest of our sample, and none of the three is radio-loud) and are found in spectra with a range of SNR$_{1700}$ (they are not found in only high-SNR spectra). We also inspected photometric light curves from the Catalina Real-Time Transient Survey (CRTS; \citealt{Drake09}) for these three candidates, and we see no unusual variability --- these quasars varied on the order of half a magnitude or less during the time spanned by our observations. 

The three acceleration/deceleration candidates are all \civ \ BALs at the higher-velocity end of the distribution of our sample; none of our acceleration candidates has $v_{\rm min}$ below 5000 \kms \ (while 62\% of our combined sample of upper limits and candidate measurements have $v_{\rm min} \leq$ 5000 \kms). However, this trend is of relatively low statistical significance. Our results also hint that the \civ \ BAL-acceleration candidates are possibly found in wider troughs than the sample of upper limits; none of our candidates has a velocity width significantly less than $\sim$4000 \kms, and the Kolmogorov-Smirnov (K-S) test yields a $p$-value of 0.025 (this means we can reject the hypothesis that the two distributions are the same at a significance level of $\geq$ 97.5\%), indicating that the difference in velocity widths between the acceleration candidates and the sample of upper limits could be significant. However, the actual significance is reduced due to the number of trials, so we consider this to be suggestive rather than compelling evidence. The equivalent widths (EWs; see Table \ref{tbl:candidates}) of the acceleration-candidate BALs do not significantly deviate statistically from those in the sample of upper limits, though we see low ($<$~10~\AA) EW values among our three candidates. 

We also searched the literature for additional spectra of our three quasars hosting acceleration candidates that might allow us to extend the time baseline covered by providing an additional, earlier epoch. J012415.53$-$003318.4 was included in the study by \cite{Gosset97}, who present spectra of quasars observed in the 1980s. If the BAL had been accelerating at our measured average acceleration (0.630~cm~s$^{-2}$), we would expect to see a shift of about 1150 \kms \ between the earlier spectrum (taken in October 1983) and our SDSS spectrum.  A comparison between the SDSS/BOSS/TDSS spectra and the prior spectrum suggests that the BAL feature in question was present when the early spectrum was taken, but we are unable to determine whether there is a velocity shift due to the low S/N and low resolution of the prior spectrum. None of the other quasars hosting acceleration candidates appears to have published spectra prior to the SDSS programs. 

\subsubsection{Other Ionization Species} 
\label{sec:otherspecies}
As discussed in the Appendix, when spectral coverage allows, we also examined the Lyman-$\alpha$/\nv, \siiv, \aliii, and \mgii \ regions of the spectra to determine if there is corresponding acceleration in other species. \cite{Filizak14} examined the variability characteristics of \civ \ BALs in conjunction with whether or not they also hosted \siiv \ and/or \aliii \ BALs and found that \civ \ troughs accompanied by \siiv \ and/or \aliii \ troughs are generally less variable than those without (as they are presumably more likely to be saturated) and that \aliii \ troughs are more variable than their \civ \ and \siiv \ counterparts. Because of these observed trends, the collective behavior of the different species can yield information on drivers of the observed trough variability. \cite{Filizak14} assigned designations to the \civ \ BAL troughs depending on which additional species are present: \civ$_{\rm 00}$ refers to \civ \ BAL troughs with no detection of BALs or mini-BALs at corresponding velocities in either \siiv \ or \aliii, \civ$_{\rm S0}$ refers to \civ \ troughs accompanied by a \siiv \ BAL at corresponding velocities but with no detection of a BAL or mini-BAL in \aliii, and \civ$_{\rm SA}$ refers to BAL troughs accompanied by both a \siiv \ and \aliii \ BAL at corresponding velocities. 

In our three candidate quasars, the Lyman-$\alpha$/\nv \ region was either not covered by the spectrograph, showed no BAL features, or had too low S/N to search for acceleration. J012415.53$-$003318.4 contains no additional BAL or mini-BAL features beyond \civ \ (and thus falls into the \civ$_{\rm 00}$ category). J013656.31$-$004623.8 has a \siiv \ BAL complex located at corresponding velocities to the \civ \ BAL acceleration candidate (falling in the \civ$_{\rm S0}$ category); however, the \siiv \ feature varies in profile shape (consistent with the trends found by \citealt{Filizak14}) and we are therefore unable to constrain acceleration of this feature. J091425.72+504854.9 hosts \siiv \ BAL features at low velocities, but none at higher velocities corresponding to the \civ \ BAL acceleration candidate trough.

\subsubsection{\civ \ Acceleration Candidates Compared to Other Studies} 
\label{sec:otherstudies}
As noted in Section 1, there are a few prior studies of individual objects that report potential acceleration of quasar absorption systems. \cite{Joshi14} report deceleration in two \civ \ BALs with magnitudes of $-$0.7~cm~s$^{-2}$ and $-2.0$~cm~s$^{-2}$; however, both of these systems show significant BAL profile variability and thus would not have been considered candidates in our study. As in our sample, we cannot rule out a velocity shift in these two variable cases, but the apparent velocity shift observed could easily be due to BAL substructure variability (e.g., ionization-driven variability or variability due to motions internal to the absorber) rather than global deceleration of the material. The previous reports of BAL acceleration by \cite{Vilkoviskij01} and \cite{Rupke02} measure $a~=$~0.035~$\pm$~0.016~cm~s$^{-2}$ and $a~=$~0.08~$\pm$~0.03~cm~s$^{-2}$, respectively, though in both cases it is difficult to assess whether or not the absorption features in question would meet our criteria for acceleration candidates. \cite{Hall07} report an acceleration of 0.154~$\pm$~0.025~cm~s$^{-2}$, and \cite{Gabel03} report a deceleration of a narrower \civ \ absorption trough of 0.1~$\pm$~0.03~cm~s$^{-2}$ and 0.25~$\pm$~0.05~cm~s$^{-2}$ (we again see some velocity-dependent profile variability in this case, but based on our criteria it appears to be a fairly convincing velocity shift). Hall et al. have since examined an additional BOSS epoch on their object, in which the variation was no longer a clear-cut case of acceleration; while they do not rule out acceleration in the first set of epochs, the results can plausibly be explained by velocity-dependent variability within the BAL.

Taking these reports at face value, the measurements by \cite{Vilkoviskij01} and \cite{Rupke02} are an order of magnitude smaller than the accelerations we are detecting in our three acceleration candidates; we measure accelerations closer to the magnitudes measured by \cite{Gabel03}, \cite{Hall07}, and \cite{Joshi14}. \cite{Vilkoviskij01} and \cite{Rupke02} both used higher-resolution spectrographs and higher-S/N spectra and are thus able to detect smaller velocity shifts than we can with the SDSS spectra. However, our use of CCF centroids to search for shifts improves our sensitivity (i.e., we are sensitive to sub-pixel shifts in the spectra, whereas previous studies generally only explored shifts in integer multiples of pixels), and we are able to obtain upper limits down to velocity shifts on par with those observed by \cite{Vilkoviskij01} and \cite{Rupke02}. 

%%%%%%%%Upper Limits %%%%%%%%%%
\subsection{Upper Limits and Constraints from Non-Detections} 
\label{sec:upperlimits} 

The majority of our targets did not show significant signatures of acceleration over the time spanned by these data; however, we can place upper limits on the magnitude of possible acceleration for many of them. We first restrict our sample to only those targets where the $p$-value (discussed in Section \ref{sec:shapetest}) is $p >$ 0.1 in the unshifted pairs of spectra, indicating a good match between the profiles and thus that the change in BAL shape and/or strength was minimal from one epoch to the next. Only 27 of the 151 total BAL complexes examined show small enough variability in shape and/or strength to allow an upper limit on the acceleration to be obtained using the entire trough.  For these troughs, we adopt the 3$\sigma$ upper limit (as characterized by the uncertainties in the velocity shift measured via the CCF using the entire trough) as upper limits on the allowed velocity shift and transform them into acceleration by dividing by the time between the epochs. We also use the absolute magnitude of the 3$\sigma$ lower limits on velocity shifts as upper limits on the magnitude of allowed deceleration. The velocity-shift upper and lower limits span the range from about half of a pixel (one pixel = 69 \kms) to $\sim$25 pixels in the worst cases, and the median 3$\sigma$ upper limit on velocity shifts is $\sim$130~\kms, which is just under the span of two pixels.

\begin{figure}
\begin{center}
\includegraphics[scale = 0.48, angle = 0, trim = 0 0 0 0, clip]{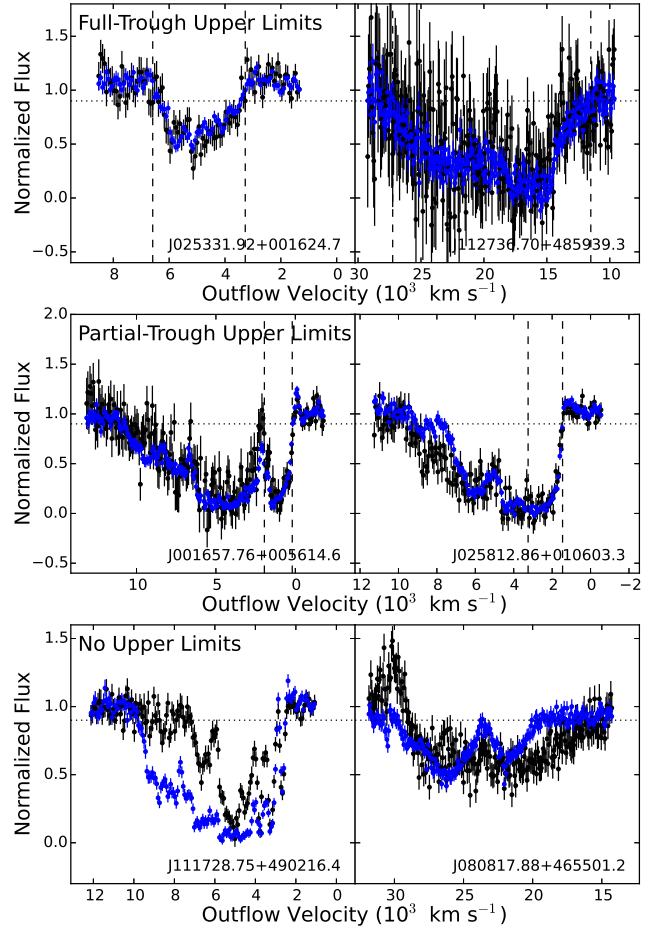} 
\caption{Examples of pairs of epochs for which we obtained upper limits from the full trough (``full-trough" upper limits; top panels), from only the nonvariable parts of the trough (``partial-trough" upper limits; middle panels), and for which we were unable to obtain upper limits due to widespread trough profile and/or strength variability (lower panels). In the top and middle panels, the ranges of the trough used in the upper-limit calculations are shown as dashed vertical lines. The horizontal dotted line represents a normalized flux of 0.9 to guide the eye. }
\label{fig:upperlim_examples}
\end{center}
\end{figure}

\begin{figure*}
\begin{center}
\includegraphics[scale = 0.48, angle = -90, trim = 0 0 0 0, clip]{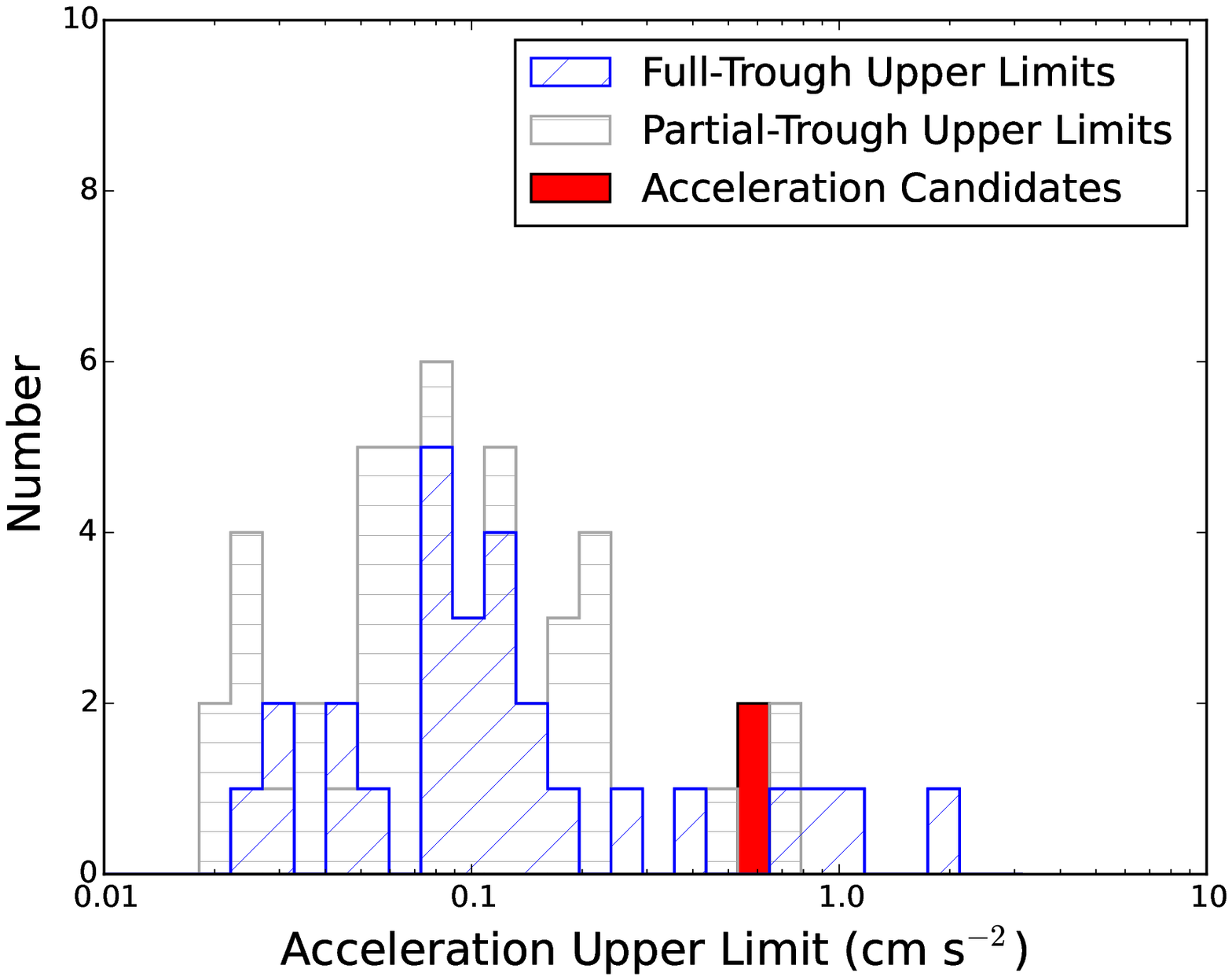} 
\includegraphics[scale = 0.48, angle = -90, trim = 0 0 0 0, clip]{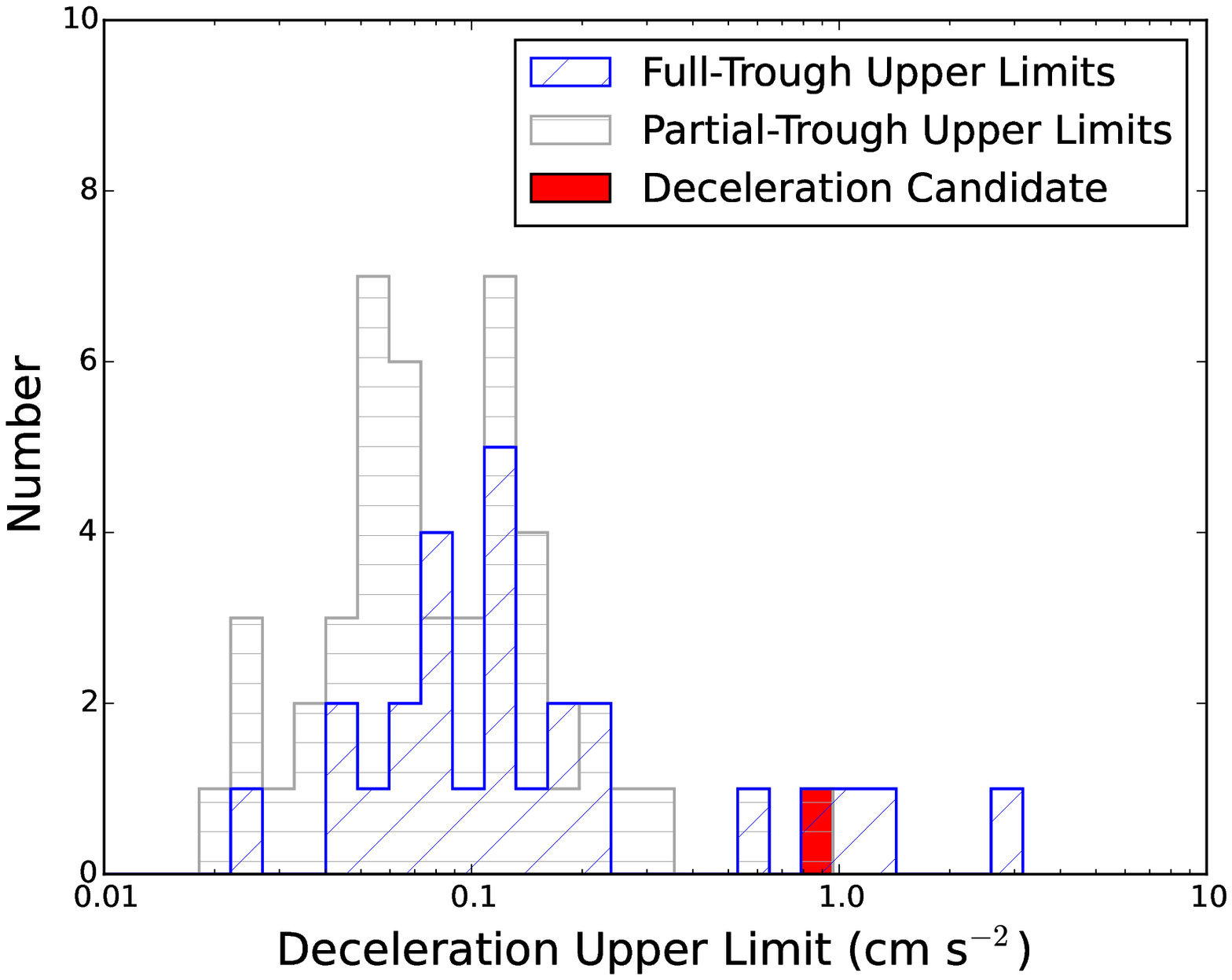} 
\caption{Left: A histogram of the magnitude of all 27 full-trough 3$\sigma$ acceleration upper limits (blue diagonal stripes) and 49 partial-trough 3$\sigma$ upper limits (gray horizontal stripes). The two histograms are overlapping, and are not additive. Right: A histogram of the magnitude of all 27 full-trough 3$\sigma$ deceleration upper limits (blue diagonal stripes) and 49 3$\sigma$ partial-trough deceleration upper limits (gray horizontal stripes). }
\label{fig:upperlim}
\end{center}
\end{figure*}

We hereafter refer to these as ``full-trough" 3$\sigma$ upper limits; these measurements are provided in the first section of Table~\ref{tbl:bal_params}. The top panels of Figure~\ref{fig:upperlim_examples} present two pairs of spectra for which we obtained full-trough upper limits, and Figure~\ref{fig:upperlim} shows histograms of the upper limits on the magnitude of acceleration and deceleration for the full-trough sample. For three quasars, the full-trough upper limit is poorly constrained due to noisy, shallow BAL profiles. We do not see any evidence that these troughs have in fact accelerated at a low significance --- the larger upper limits are simply due to the lower-quality data in these three cases. 

The remaining 121 troughs (neglecting the 3 troughs that are real acceleration/deceleration candidates) showed significant variability in trough shape and/or strength between the SDSS and TDSS observations in at least part of the trough.  We know from previous work (e.g., \citealt{Gibson08}) that variations tend to occur in only portions of troughs. Indeed, in our targets we see in many cases that only part of the BAL trough varied; in such cases, meaningful upper limits can still be derived by examining the region near \vmin or $v_{\rm max}$ if the BAL flux in that region did not vary significantly, as a variable trough would still show an increase or decrease in velocity at the upper and lower limits of the trough if it had accelerated. We inspected each set of observations and identified those troughs that only varied in part of the trough and for which we can still derive meaningful upper limits to their acceleration. For 55 of the 121 cases where the BAL varied, the variability was isolated to only part of the trough --- we identified these cases by eye. The middle panels of Figure~\ref{fig:upperlim_examples} display two examples of such cases. The other 66 BAL troughs showed widespread variability across the trough in strength and/or profile shape, so we were unable to obtain upper limits for acceleration. Two examples of such cases are presented in the bottom panels of Figure~\ref{fig:upperlim_examples}. 

To obtain upper limits for the 55 BALs that only showed partial variability, we first isolated the nonvariable part of the trough complex near \vmin or $v_{\rm max}$. In most cases, the BAL was steepest and least variable at the low-velocity end; this result is consistent with prior reports that shallower parts of troughs are more variable (e.g., \citealt{Capellupo11}) and/or that high-velocity portions of \civ \ BAL troughs tend to be more variable (\citealt{Filizak14}). We cropped the spectra manually to include only the non-variable portion of the trough; the proportion of the BAL used in the analysis varied from object to object, but we typically included about a third to half of the original BAL trough. These ``partial-trough" spectra were re-run through the CCF analysis to search for velocity shifts. There were six cases where we detected velocity shifts (at 3$\sigma$ significance) using partial troughs; however, in all cases, the $\chi^2$ test failed and visual inspection indicates that the shift was caused by a widening or narrowing of the trough profile. We discard these six measurements from our upper-limits sample. 

In the 49 remaining partial troughs, we find no significant detections of acceleration, so we adopt the 3$\sigma$ upper uncertainties in the measured velocity shift to calculate our formal upper limits on BAL acceleration and the 3$\sigma$ lower limits to calculate formal upper limits on BAL deceleration.The measured upper limits on the acceleration and deceleration for 49 partial BAL troughs are given in the second section of Table \ref{tbl:bal_params}. Histograms of these partial-trough upper limits on acceleration and deceleration are presented in Figure~\ref{fig:upperlim}. The median velocity-shift upper limit for the partial-trough sample was $\sim$100~\kms, which spans almost 1.5 pixels, and the velocity-shift upper limits range from about one tenth of a pixel in the highest-quality spectra $\sim$15 pixels in the lowest-quality spectra. These statistics suggest that the upper limits for the partial-trough sample are overall a bit smaller than those for the full-trough sample, which is due partially to higher S/N in the partial-trough sample (the median SNR$_{1700}$ of the full-trough sample is about 13.6, and the median SNR$_{1700}$ of the partial-trough sample is 16.6) and also because there is a higher proportion of steep BAL profiles in the partial-trough sample than in the full-trough sample. These effects reduce the uncertainties in the measured velocity shifts and thus yield tighter constraints. 

Our measured upper limits on acceleration for the entire sample of full-trough and partial-trough measurements range from 0.002~cm~s$^{-2}$ to 1.8~cm~s$^{-2}$ overall, with a median acceleration upper limit of about 0.1~cm~s$^{-2}$. We were able to set upper limits on velocity shifts in 90\% of the full-trough and partial-trough BALs to magnitudes of less than 3\% of the mean velocity of the BAL; the few exceptions are in noisier, shallower BAL troughs. As discussed in Section \ref{sec:ccfsec}, the minimum velocity shift that we are sensitive to in our study depends on several factors (the S/N of the spectra, the shape of the BAL features, the number of pixels used in the CCF measurement, and the spectral resolution). These factors lead to a range in upper-limit constraints for different quasars. 

\section{DISCUSSION AND FUTURE WORK} 
\label{sec:discussion}

We have performed an investigation of \civ \ BALs in 140 quasars having three epochs of spectroscopy from SDSS, BOSS, and TDSS. In this sample, we examined 151 distinct \civ \ BALs in 121 different quasars, yielding the tightest observational constraints on BAL acceleration to date. The main observational results of this study can be summarized as follows: 
\begin{enumerate}  
\item We measured significant acceleration or deceleration signatures in three BAL troughs, two of which we consider solid candidates for acceleration. In addition, we obtained robust upper limits on acceleration for 76 BALs that varied minimally in strength and/or profile shape. The remaining 72 BAL troughs showed significant variability in strength and/or profile shape, and thus we were unable to obtain useful constraints for them (see Section~\ref{sec:accelsearch}). 
\item The magnitudes of positive acceleration measured in the two most solid acceleration candidates are 0.63$^{+0.14}_{-0.12}$~cm~s$^{-2}$ and 0.54$\pm 0.04$~cm~s$^{-2}$ (Table~\ref{tbl:candidates}) over rest-frame time frames of 3.7 and 5.2 years, respectively. In both cases, there is evidence that the magnitude of the acceleration decreases over time (the magnitude is smaller between Epochs 2 and 3 than it is between Epochs 1 and 2; see Section \ref{sec:civaccel}).  Additional epochs of data for these targets would be of particular interest.  
\item One case presents possible evidence for deceleration (a decrease in outflow velocity) at a magnitude of 1.14$\pm0.20$~cm~s$^{-2}$ over a period of 3.4 years in the quasar rest frame between Epochs 1 and 2 (Section \ref{sec:civaccel}). However, we are only able to establish an upper limit to the acceleration/deceleration between Epochs 2 and 3 due to significant weakening of the trough over this time period. 
\item We were able to place upper limits on acceleration/deceleration for the majority of the sample to be less than 0.2~cm~s$^{-2}$ (Section~\ref{sec:upperlimits} and Figure~\ref{fig:upperlim}). The magnitude of the upper limits depends on the properties of the BAL (e.g., width, steepness of onset) as well as the quality of the spectra (e.g., S/N). 
\item The tight constraints on acceleration/deceleration for most of our BALs indicate that the majority are stable in velocity to within 3\% of the mean velocity of the trough, even over timescales of a few years in the quasar rest frame  (Section \ref{sec:upperlimits}). 
\end{enumerate} 

Our study includes the largest sample of BALs and utilizes more rigorous methodology for setting acceleration constraints than the previous relevant studies of BAL variability (see Sections \ref{sec:introduction} and \ref{sec:otherstudies}). Prior studies reporting possible acceleration candidates were focused on individual objects rather than large samples. The previous studies listed in Table \ref{tbl:previous_studies} do not provide quantitative upper limits on acceleration, with the exception of \cite{Gibson08} who provide upper limits on acceleration for seven \civ \ BALs encountered in their general exploration of BAL variability. However, to obtain these upper limits, they use only very small portions of the trough and assume a maximum velocity shift of one pixel to derive their upper limits. As discussed in Section~\ref{sec:ccfsec}, the major advantages of our CCF/$\chi^2$ methodology are the ability to resolve ambiguities between velocity-dependent variability and acceleration by using the entire trough (or at least a significant fraction of it) and the rigorous quantification of uncertainties, which allows derivation of robust upper limits on velocity shifts and acceleration/deceleration. From our sample of three acceleration candidates and 76 upper limits, we are able below to examine possible models for the observed acceleration or deceleration as well as examine the occurrence rate of acceleration and deceleration in BAL systems. %As noted above, the velocity sampling of the SDSS, BOSS, and TDSS spectra is about 69 \kms \ per pixel in the \civ \ region.  

\subsection{Disk-Wind Models and BAL Acceleration} 
\label{sec:windmodels} 
For the two solid acceleration candidates, we can investigate whether our observations are plausibly consistent with disk-wind models describing the production and behavior of BALs in quasars. In a typical disk-wind model, the outflow is accelerated by incident ionizing radiation; therefore, for a source with a constant luminosity, one could plausibly expect to see an increase in the velocity of a BAL over time. 
%\subsubsection{Acceleration by Radiation Pressure}  
Adopting the model of \cite{Murray95}, the radial velocity of a radiatively accelerated wind is approximately
\begin{equation} 
\label{eq:vr}
v(r)~=~v_\infty~(1~-~r_L/r)^{1.15}
\end{equation} 
where $v_{\infty}$ is the terminal outflow velocity, $r_L$ is the launching radius, $r$ is the current radius, and $v(r)$ is the current velocity of the gas (See Section 4.1 of \citealt{Murray97}). 
Equation \ref{eq:vr} yields an acceleration of 
\begin{equation} % = \frac{dv}{dr}\frac{dr}{dt}
\label{eq:accel}
a(r) = \frac{dv}{dt} = v\frac{dv}{dr}
= 1.15\frac{v_\infty^2 r_L}{r^2}\left(1-\frac{r_L}{r}\right)^{1.30}.
\end{equation}

We assume acceleration due to radiation pressure (e.g., \citealt{Murray95}; \citealt{Baskin14}), which means $v_\infty = F \sqrt{GM/r_L}$, where $F$ (the ratio between $v_{\infty}$ and $v_{\rm circ}$ at the launching radius $r_L$) is on the order of a few ($1.5 < F < 3.5$; \citealt{Murray95}; \citealt{Rogerson15}, although \citealt{Risaliti10} find values as high as 10$-$20). Rearranging this equation,   
$r_L~=~F^2GM/{v_\infty}^2 $, and we can eliminate $r_L$ in Equation \ref{eq:accel} by substitution, such that: 
\begin{equation} 
\label{eq:accel_vonly} 
a(r) = 1.15\frac{F^2GM}{r^2}(1-\frac{F^2GM}{rv_\infty^2})^{1.30}.
\end{equation} 

We have measurements of $a$, $M$, and the observed velocity $v$ of the BAL for our two acceleration candidates; $F$ and $r$ are model parameters. This particular model will be successful if we can simultaneously satisfy Equations~\ref{eq:vr} and \ref{eq:accel_vonly} with our measurements using reasonable values of $F$ and $r$. We use J012415.53$-$003318.4 to demonstrate. For this quasar, we have $M_{\rm BH}~\sim~2.0~\times~10^9$~\Msun \ (\citealt{Shen11}).  If we consider the average acceleration and observed velocities between Epochs 1 and 3 ($a$~=~0.630~cm~s$^{-2}$ and spanning $\sim$16,500~--~20,500~\kms \ in velocity), the above equations are satisfied if we use $F~=~$3 and $r/r_L = 5$ (i.e., we are observing the wind at about five times the launching radius). In this case, the model predicts a launching radius $r_L \sim 3.6 \times 10^{17}$ cm ($\sim 0.12$ pc) and a terminal velocity $v_{\infty}~\sim$~25500 \kms. This means we are observing the BAL at a radius of $r_{\rm obs} \sim$ 0.6 pc. We can similarly satisfy the above equations using the measured acceleration between Epochs 1 and 2 or Epochs 2 and 3 individually (i.e., treating each as a separate, isolated case). Thus, for a single acceleration measurement, the above model is viable. 

However, for our acceleration candidates, the acceleration is not constant throughout time --- we have measurements of the average acceleration between Epochs 1 and 2 ($a_{1-2}$) and Epochs 2 and 3 ($a_{2-3}$), which are not consistent with one another for either of our targets. The rate of change of the acceleration is known as the jerk --- a successful model must also be able to reproduce this quantity to remain viable. 
For the BAL in J012415.53$-$003318.4, we observe $a_{1-2}~=~0.90^{+0.21}_{-0.32}$~cm~s$^{-2}$ and $a_{2-3}~=~0.37^{+0.17}_{-0.10}$~cm~s$^{-2}$. Each of these acceleration measurements is actually the average acceleration over the pair of epochs examined, so the midpoint between each pair of epochs is the best characterization of the time we observed each acceleration magnitude. The rest-frame time difference between these two midpoints is 1.84 years. Dividing the change in acceleration over this period ($-0.53^{+0.27}_{-0.33}$~cm~s$^{-2}$) by the rest-frame time difference between these measured acceleration magnitudes yields an average jerk of $j~=~-0.29$~cm~s$^{-2}$~yr$^{-1}$. If the jerk remains constant, we would expect this BAL to reach an acceleration of zero in about 1.3 years in the quasar rest frame (about 4 years in the observed frame). Hence, additional observations over the next decade will be particularly instructive in investigating the evolution of the acceleration with time in this target. 

In order for our adopted disk-wind model to be viable, we must be able to reproduce simultaneously {\it both} acceleration measurements via the above equations with the same values of $F$ and $r_L$ (effectively reproducing the measured jerk), while also obtaining reasonable predicted values for the observed velocities. We are unable to match the observed rapid change in acceleration while also matching the observed velocity shifts of the BAL --- setting our observed acceleration values and fixing $F$ and $r_L$ to satisfy the equations for $a_{1-2}$ causes the model to over-predict the velocity shift by about a factor of five when these values are applied with $a_{2-3}$. We conclude that the model cannot accommodate such a large jerk and thus cannot successfully describe our observations for this candidate. We can do the same calculations for J013656.31$-$004623.8, and we are again able to produce the individual acceleration measurements for this target under reasonable assumptions, but the disk-wind model is unable to accommodate the large magnitude of the jerk. We thus conclude that our adopted wind model is insufficient to describe our results --- different and/or more complicated models are likely required to explain correctly the observed changes in acceleration. 

We note further that for a wind driven by a source of variable luminosity, changes in the line-driving flux would cause variations in the magnitude of the observed acceleration --- thus, in the context of this model, the observed jerk could be produced by changes in the flux incident on the outflow. However, as noted in Section~\ref{sec:civaccel}, we observe no extraordinary photometric variability of the quasars hosting our acceleration candidates. While the observed photometric flux is not a perfect tracer of the line-driving flux, the lack of exceptional optical variability of these quasars supports the idea that the acceleration and the jerk are due to causes not well-described by the disk-wind model. 

Gas in a standing-flow pattern (as discussed in Section~1) across and along our line of sight might show apparent acceleration or deceleration if the launching radius $r_L$ of the outflow were variable, e.g., due to line-driving luminosity changes of the quasar. This would result in a change in $v_\infty=F\sqrt{GM/r_L}$ and thus $v(r)$ at fixed $r$ (although it is possible that such a change would also cause the profile of the BAL to change, which we select against in our study). The low incidence of observed acceleration/deceleration in our sample and the lack of extraordinary photometric variability in our acceleration candidates compared to the rest of the sample together suggest that variations in the launching radius are unlikely to be common. However, whether such a model can match all of our observations in detail will require further study. 

Since this idea has not yet been explored in the literature, we briefly consider the effect that a variable $r_L$ could have on our BALs. For example, an outward migration of $r_L$ across the disk at a speed of 10\% of the circular velocity at that radius would yield an observed deceleration of  $-0.22$~cm~s$^{-2}$ in J012415.53$-$003318.4. This deceleration would occur {\it in addition} to the radiative acceleration discussed above, and thus could explain some of the jerk observed between the SDSS--BOSS and BOSS--TDSS epochs in that object. The same applies to J013656.31$-$004623.8. However, explaining the deceleration observed in J091425.72+504854.9 would require a faster change in the launching radius, as this mechanism would have to produce deceleration large enough in magnitude to overcome the positive acceleration caused by radiation. The range of radial speeds with which a launching radius can plausibly change would be a useful quantity to determine in future theoretical studies of wind launching.

\subsection{Geometric Effects From Disk Rotation} 
\label{sec:diskrotation} 
Apparent acceleration or deceleration of a BAL due to geometric projection effects is another possible scenario. Assuming that BAL material is launched from a rotating disk, one would expect the disk wind to continue to rotate as it travels outward; such rotation could produce either an acceleration or deceleration signature, depending on the system geometry (e.g., see Section 5.3 of \citealt{Hall13}). To assess the timescale over which this may be observed, we consider the rotation period of material in a disk. BAL material is thought to be launched at radii of $\approx$1000 times the Schwarzschild radius of the black hole (e.g., \citealt{Murray95}; \citealt{Gibson08b}), which for a 10$^{9}$ \Msun \ black hole (similar to those hosted by the quasars of our two acceleration candidates), would be on the order of 10$^{17}$ cm. The orbital period of material at this radius is $\approx$17 years. Thus, over a timespan of 2--5 years in the quasar rest frame (which is the typical range covered by our observations, as demonstrated by Figure~\ref{fig:timescales}), material located at the launching radius of a rotating disk would be expected to complete about a tenth to a third of a full rotation. This would be a substantial change in the angle of the material with respect to our line-of-sight; thus it would be surprising to see no acceleration or deceleration signatures of material over this timespan if these BALs were launched at radii $\sim$10$^{17}$ cm. The observed level of velocity stability (to within 3\% of the mean velocity of the BAL trough) would require remarkable azimuthal symmetry of the disk wind. 

On the other hand, BALs may be typically observed at larger distances than their launching radii ---  for example, consider gas located at radii on the order of 10$^{18}$ cm.  The rotation period of disk material at this radius is on the order of 500 years. As the outflow now would not have rotated substantially with respect to our line-of-sight over the period it was observed, it is thus less likely that we would observe acceleration or deceleration due to rotation of the BAL wind about the disk (if we assume that disk rotation is the cause of acceleration or deceleration).  

Beyond causing acceleration, we consider the possibility that rotational effects could also be responsible for the observed jerk in our two acceleration candidates. The case of a rotating disk and the contribution of the rotational component of the wind's velocity to our observed line-of-sight velocity was explored by \cite{Hall02}  --- see their Section 6.5.2. Unfortunately, the exact contribution of the rotational velocity component depends heavily on the details of the system, and we are unable to consider exhaustively all of the possible scenarios to determine whether the observed acceleration and jerk can be plausibly explained via rotation of the gas with the disk. However, we explored a number of possible scenarios, and find that it is difficult to explain both the magnitudes of the acceleration and jerk simultaneously when considering solely the rotation of the disk. When matching our observed acceleration, the magnitude of the predicted jerk is too small by several factors (and sometimes by a full order of magnitude, depending on our assumptions). We thus find it likely that the rotation of the disk cannot be the only cause of the observed BAL acceleration and jerk. 

\subsection{BAL Deceleration} 
\label{sec:deceleration} 

While two of our candidates have accelerated toward higher velocities, we see possible BAL deceleration in one object (J091425.72+504854.9). Although it is not one of our most solid cases, since we only have upper limits on acceleration/deceleration between Epochs 2 and 3 (and we even see marginal evidence that the BAL may have increased in velocity between these two epochs), the deceleration signature between the first two epochs is fairly convincing. Considering our model in Section~\ref{sec:windmodels}, we would not expect to observe deceleration of the BAL due to changes in line-driving flux; such variability would only cause decreases in the magnitude of the acceleration (i.e., a negative jerk). To explain apparent deceleration of a BAL requires alternate scenarios. 

Several studies have suggested that low-ionization BAL absorbers can be located at large distances (100--1000~pc-scale) from the central source and are energetic enough to provide significant feedback to the host galaxy (e.g., \citealt{Moe09}; \citealt{Dunn10}; \citealt{Leighly14}). Even more recently, evidence has emerged that some high-ionization BAL absorbers, which are more common in quasars than the low-ionization BALs, can also be found at large distances and provide feedback (e.g., \citealt{Arav13}; \citealt{Borguet13}; \citealt{Chamberlain15}), though this is not seen in all cases (e.g., \citealt{Chamberlain15b}). Such feedback interactions with ambient material could plausibly cause BAL winds to decelerate. The effects of the wind on the interstellar medium and on the host galaxy have been discussed extensively in the context of quasar feedback (e.g., \citealt{FaucherGiguere12}), and observations by \cite{Leighly14} present evidence for the acceleration/compression of the ambient material by a BAL outflow. However, to our knowledge, past theoretical work has not discussed quantitatively the effect of such an interaction on the velocity of the BAL outflow itself (Faucher-Gigu\`{e}re 2016, private communication). 

The majority of our BALs do not show significant evidence for deceleration, and our upper limits are tight in many cases. If we are observing these BALs when they are located at small radii from the central source, they may not yet have traveled sufficiently far to encounter ambient material and thus are not decelerating. Alternatively, as discussed in Section \ref{sec:introduction}, this lack of observed deceleration could indicate that the wind disperses and is optically thin before it reaches any material. 

\subsection{BAL Acceleration Occurrence Rates}
\label{sec:rates}
We measured the velocity shifts and inferred acceleration/deceleration of three BAL troughs in the sample of 151 \civ \ BAL-trough complexes (two of which we consider solid candidates for acceleration) and obtained upper limits on the acceleration and deceleration in 76 additional BAL-trough complexes. Because of the difficulty of disentangling acceleration/deceleration from BAL strength and profile variability (e.g., due to ionization-driven variability or motions internal to the absorber), we cannot demonstrate that the other 72 \civ \ BAL complexes did not accelerate; instead, they are considered cases where we would not be able to detect actual acceleration even if it had occurred. Nevertheless, the 79 targets for which we have obtained measurements or upper limits on the acceleration can be used to estimate the incidence of BAL acceleration.  

Out of 79 measurements, acceleration above a magnitude of 0.5~cm~s$^{-2}$ could be present in up to eight BALs: our two acceleration candidates as well as six weak upper limits. The fractional range is thus $\sim$3--10\%. The lower end of this range is calculated by discarding the largest upper limits, as they are most likely due entirely to poor-quality data rather than the potential for acceleration. When considering deceleration, we have one detection and seven upper limits above a magnitude of 0.5~cm~s$^{-2}$ (though the upper limits are again most likely due to poor-quality spectra rather than an increased potential of acceleration). The possible deceleration rate is thus 1--10\% in our sample. We note that these occurrence rates are likely lower limits, as our rates are subject to the limitations of our analysis methods (discussed, e.g., in Section \ref{sec:shapetest}). Our sample was drawn from a classical sample of luminous BAL quasars and spans a range of redshift, luminosity, and \mbh (see Figure \ref{fig:mbhz}), so these results can thus be applied to populations of typical BAL quasars. 

These low rates could have a number of different explanations. For example, consider gas in a standing-flow pattern as discussed in Sections~\ref{sec:windmodels} and \ref{sec:diskrotation}. If the BALs are observed at small radii, it would require remarkable azimuthal symmetry of the disk wind to produce a BAL that is stable in velocity to the degree observed over rest-frame timescales of several years (i.e., one would expect to see acceleration/deceleration due to rotation of the disk). In order for this issue to be alleviated, the gas would have to be quite distant from the central black hole --- thus our low incidence of acceleration/deceleration could indicate that the absorbers are located at larger distances from the central source. An alternate explanation could be that the gas is slowed by drag in the surrounding medium such that it retains a constant velocity (although this latter situation would require rather contrived fine tuning to balance the radiative force and the drag force). 

The low incidence of acceleration could also be due to observational biases: If BALs that are highly variable in depth and/or shape also tend to accelerate more often, we will underestimate the rate of acceleration, since we eliminate these 72 objects from our analysis. However, our data show no signs of such a bias --- our partial-trough sample contains variable BALs, yet we see no evidence that our acceleration constraints are correlated with the amount of profile variability. We thus conclude that this possible bias is unlikely to apply. 

\subsection{Future Work} 

The parent sample of BAL quasars used in this study constitutes by far the largest sample to date of multi-epoch BAL quasar spectra probing multi-year timescales. With 151 BAL troughs in this quasar sample, we have shown that detecting BAL acceleration/deceleration is difficult, and clear cases of acceleration are relatively rare; to use observations of BAL acceleration to constrain quasar wind/outflow models effectively requires large numbers of targets. The parent sample of BAL quasars from this study continues to be observed by TDSS, so over the next few years the sample of quasars eligible for a subsequent study of acceleration will greatly increase; by the conclusion of TDSS observations, we expect to have three epochs of spectra for $\sim$1600 BAL quasars, expanding the size of our sample by nearly an order of magnitude. This much larger sample of quasars with three epochs of spectra over longer timescales will provide improved statistical constraints on the occurrence of acceleration in BAL quasars and should yield more acceleration measurements for comparison with outflow models. Given the fraction of acceleration/deceleration candidates found here (3 candidates out of 140 quasars), we expect to find $\sim$35 acceleration candidates once the entire 1600-quasar sample has been observed by TDSS. As the time between the BOSS program observations of this sample and the incoming TDSS observations grows, the additional observations will cover longer time baselines and help determine whether the low detection rate is due in part to inadequate time spans between the current BOSS vs. TDSS observations.    

At least two of the studies that found acceleration in single-object investigations (\citealt{Vilkoviskij01}; \citealt{Rupke02}) used higher-resolution spectrographs and obtained higher-S/N spectra than for our sample. While our data quality was sufficient to find a few cases, a large-sample search using high-resolution, high-S/N observations would yield even tighter constraints on BAL acceleration. Targeting our acceleration candidates at high resolution and high S/N could also provide more solid evidence for or against the possibility of acceleration as well as provide additional epochs with which to explore further trends in the magnitude of acceleration or deceleration over time. We have shown above that BAL acceleration can be observed and constrained --- future theoretical investigations that yield quantitative predictions for the magnitude of BAL acceleration would thus be advantageous, as our acceleration measurements could then be used to evaluate current quasar outflow models and further investigate the connection between outflows and their host galaxies. 

 \acknowledgments 
The authors thank J.~A.~Rogerson for providing code upon which software used in our analysis was based. We also thank our anonymous referee for useful suggestions on improving our work. We thank K.~D.~Denney and B.~M.~Peterson for useful discussions on cross-correlation methods, as well as C.~Faucher-Gigu{\`e}re and D.~Proga for discussions of acceleration/deceleration in the context of outflow models. We thank E.~Gosset for providing us with a higher-resolution plot of an earlier spectrum of one of our acceleration candidates. CJG and WNB acknowledge support from NSF grant AST-1108604 and the V.M. Willaman Endowment. PBH is supported by NSERC. JRT acknowledges support from NASA through Hubble Fellowship grant \#51330, awarded by the Space Telescope Science Institute, which is operated by the Association of Universities for Research in Astronomy, Inc., for NASA under contract NAS 5-26555. NFA acknowledges support from TUBITAK (115F037). 

Funding for the Sloan Digital Sky Survey IV has been provided by
the Alfred P. Sloan Foundation, the U.S. Department of Energy Office of
Science, and the Participating Institutions. SDSS-IV acknowledges
support and resources from the Center for High-Performance Computing at
the University of Utah. The SDSS web site is www.sdss.org.

SDSS-IV is managed by the Astrophysical Research Consortium for the 
Participating Institutions of the SDSS Collaboration including the 
Brazilian Participation Group, the Carnegie Institution for Science, 
Carnegie Mellon University, the Chilean Participation Group, the French Participation Group, Harvard-Smithsonian Center for Astrophysics, 
Instituto de Astrof\'isica de Canarias, The Johns Hopkins University, 
Kavli Institute for the Physics and Mathematics of the Universe (IPMU) / 
University of Tokyo, Lawrence Berkeley National Laboratory, 
Leibniz Institut f\"ur Astrophysik Potsdam (AIP),  
Max-Planck-Institut f\"ur Astronomie (MPIA Heidelberg), 
Max-Planck-Institut f\"ur Astrophysik (MPA Garching), 
Max-Planck-Institut f\"ur Extraterrestrische Physik (MPE), 
National Astronomical Observatory of China, New Mexico State University, 
New York University, University of Notre Dame, 
Observat\'ario Nacional / MCTI, The Ohio State University, 
Pennsylvania State University, Shanghai Astronomical Observatory, 
United Kingdom Participation Group,
Universidad Nacional Aut\'onoma de M\'exico, University of Arizona, 
University of Colorado Boulder, University of Oxford, University of Portsmouth, 
University of Utah, University of Virginia, University of Washington, University of Wisconsin, 
Vanderbilt University, and Yale University.

%%%%%%%%

\clearpage 
%%%%%%%%%APPENDIX 
\begin{appendix}  
\label{appendixb}

Our CCF and $\chi^2$ analyses discussed in Section \ref{sec:accelsearch} yielded three acceleration candidates. We here discuss each of these candidates in detail and evaluate the likelihood that the observed variability is due to acceleration of the \civ \ BAL. In addition, we search for and consider additional absorption species in these candidates. 
\\
%%%%%%%%%%%%%%%%%%%%%%%%%
%Number 20, J012415.53$-$003318.4, J0124
%%%%%%%%%%%%%%%%%%%%%%%%%%%
\subsubsection{SDSS\,J012415.53$-$003318.4} 
\begin{figure*}
\begin{center}
\includegraphics[scale = 0.32, angle = -90, trim = 20 0 0 0, clip]{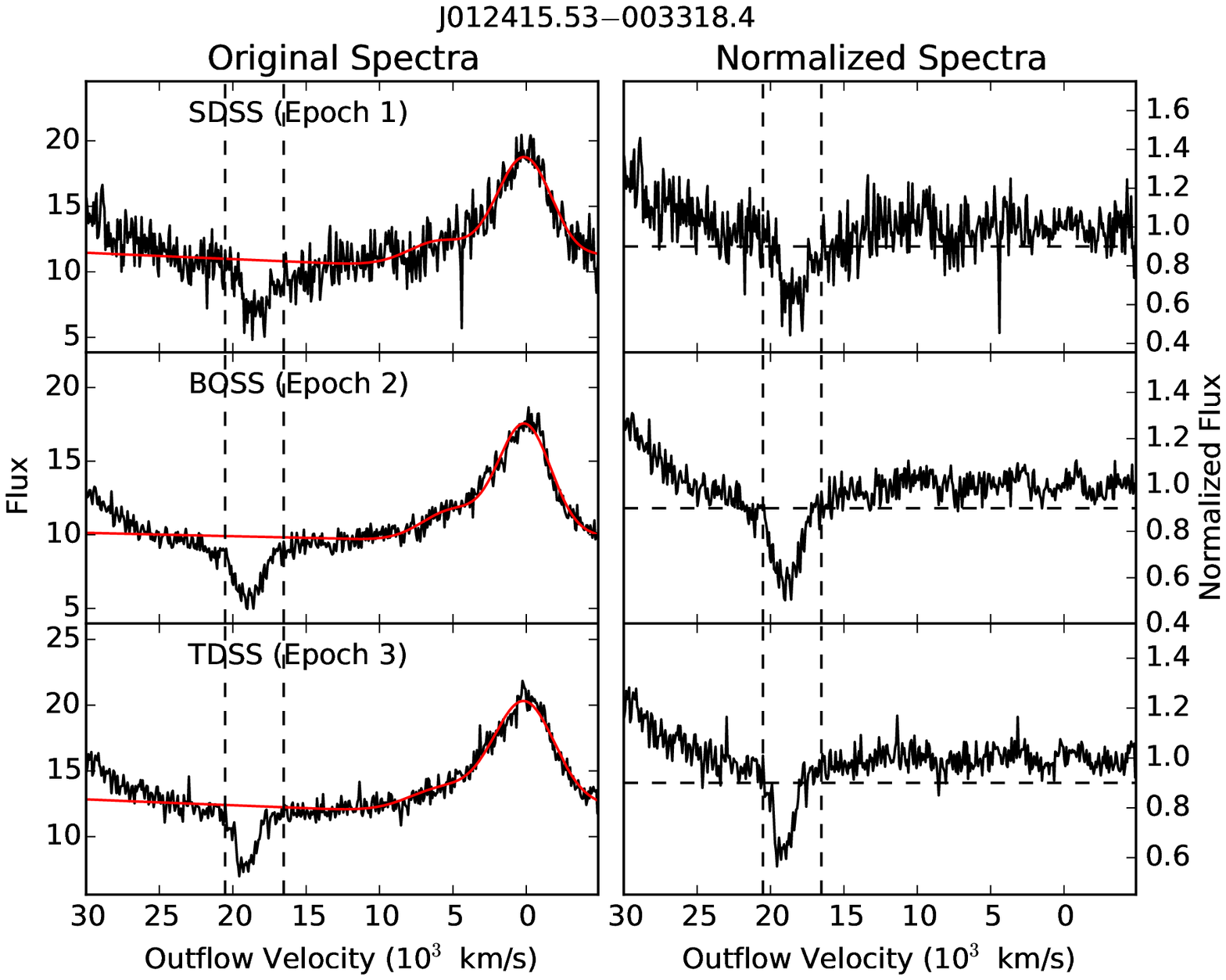} 
\includegraphics[scale = 0.32, angle = -90, trim = 0 0 0 0, clip]{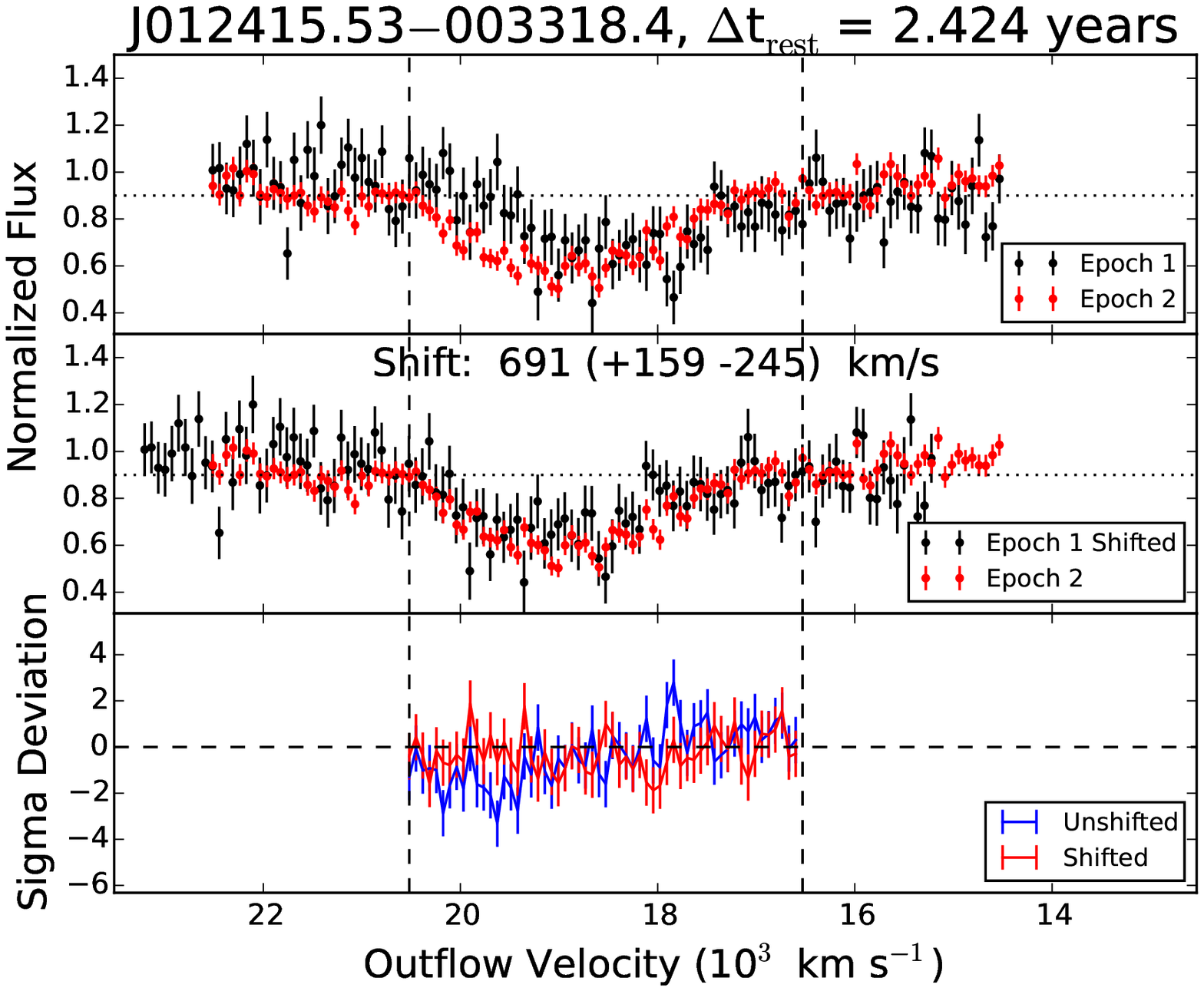} 
\includegraphics[scale = 0.32, angle = -90, trim = 0 0 0 0, clip]{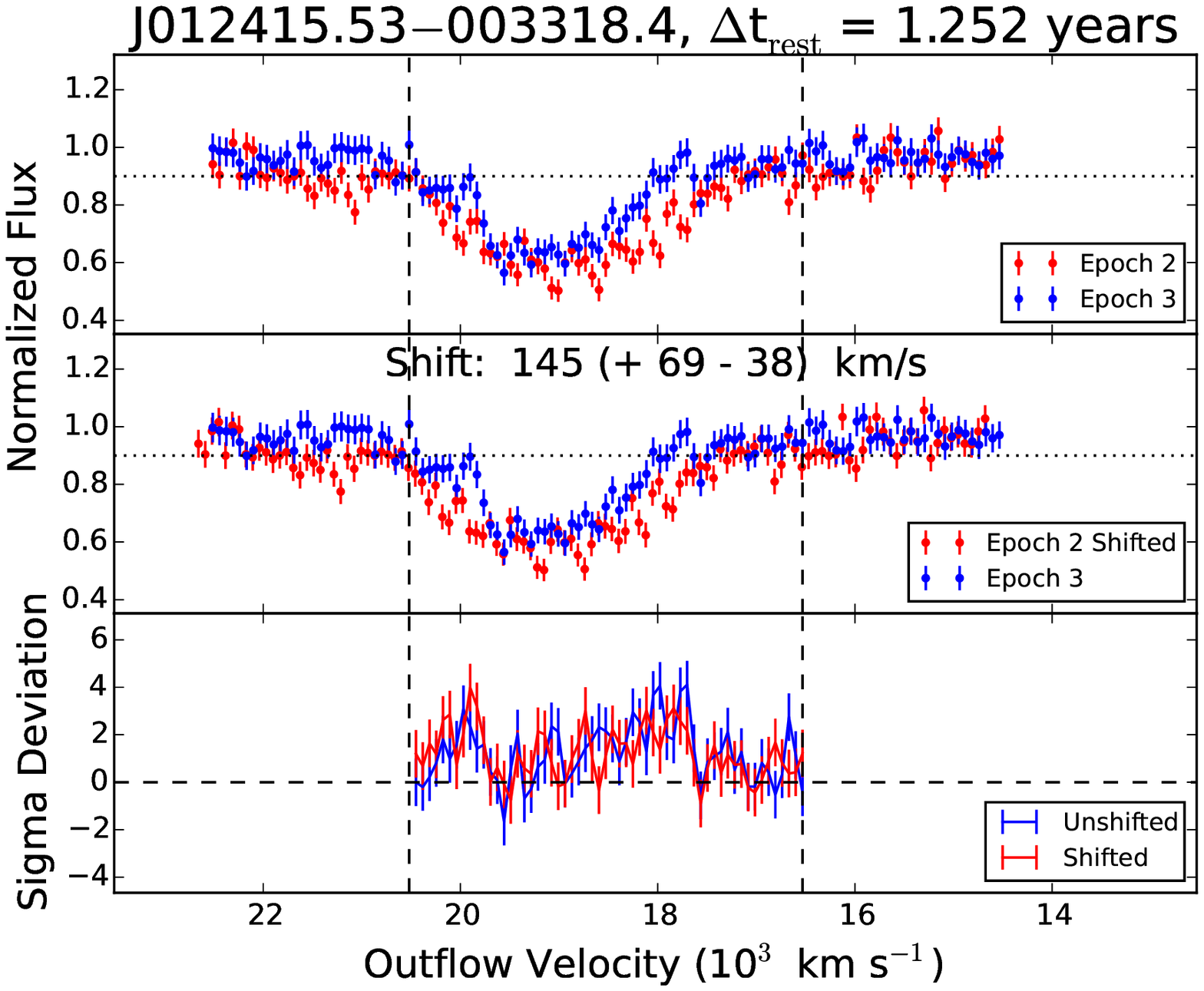} 
\caption{Spectra of the SDSS\,J012415.53$-$003318.4 \civ \ BAL. The left panels show the original spectra, the associated continuum plus \civ \ emission-line fits, and the normalized spectra; vertical dashed lines indicate the velocity range of the BAL in question. The middle and right panels present the pairs of spectra for this target. In all cases, the first epoch (SDSS) is shown in black, the second epoch (BOSS) is shown in red, and the third epoch (TDSS) is shown in blue. The middle subpanel of each panel displays the two spectra with the earlier epoch shifted by the measured velocity shift from the CCF analysis. The bottom subpanel in the middle and right panels shows the deviation (in sigma) between the spectra before and after the shift in red and blue, respectively. Despite the narrowing of the feature between Epochs 2 and 3, this BAL is one of our most solid cases of acceleration. }
\label{fig:accel2}
\end{center}
\end{figure*}

SDSS\,J012415.53$-$003318.4 (Figure~\ref{fig:accel2}) is one of our two acceleration candidates; the comparisons between Epochs 1 and 2 (middle panel of Figure~\ref{fig:accel2}) and Epochs 1 and 3 (Figure \ref{fig:all_candidates}) formally meet our requirements for acceleration candidates, although Epochs 2 and 3 (right panel of Figure \ref{fig:accel2}) do not. In addition, the lower panels of the comparison plots, showing the sigma deviation between the two spectra, reveal a trend expected from a velocity shift; the ratio of the deviation at the higher-velocity end of the BAL is systematically negative, while at the lower-velocity end of the BAL it is systematically positive. This offset is remedied by applying the velocity shift to the first epoch --- the red error bars show the sigma deviation after the shift is applied; the velocity shift removes the trend. Unfortunately, Epochs 2 and 3 (right panel of Figure \ref{fig:accel2}) do not formally meet the $\chi^2$ requirements to be considered unambiguous acceleration because the trough visibly narrowed between these two epochs, resulting in a statistically unacceptable match between the two spectra. This was not a problem in the cases involving Epoch 1 because of its lower S/N --- this situation causes even the narrowed profile in Epoch 3 to still be considered an acceptable fit. We thus consider the acceleration measurement between Epochs 2 and 3 to be relatively uncertain; while the narrowed trough does appear to have shifted and is thus plausibly accelerating, velocity-dependent variability could cause the observed changes as well. If we do treat the case of Epochs 2 and 3 as actual acceleration, the average measured acceleration in this BAL is not constant over time; the acceleration measured between Epochs 2 and 3 is less than half that between the first and second epochs (see Table~\ref{tbl:candidates}). 
%This is consistent with observations by \cite{Gabel03}, who see the magnitude of acceleration vary in their outflow as well. **The Gabel et al object is a narrower outflow, so I need to check and make sure it's worth comparing things to the Gabel et al. measurement.} %I think it was them but I'm not positive. Double-check somewhere. 

We searched the rest of the spectrum of this target for additional BALs of different species (e.g., \nv, \siiv, \mgii, \aliii) to see if our observed velocity shifts are seen in these other species as well. Unfortunately, there are no additional BALs in this target for us to examine, although there is a lower-velocity narrow absorption line (NAL) visible in the \siiv \ region. Additionally, there are some absorption features visible blueward of the Ly$\alpha$ emission line, one of which is likely a \nv \ BAL or mini-BAL that is at a similar outflow velocity as the \civ \ BAL. However, this region of the spectrum is not covered by the first epoch (SDSS) and the S/N is such that we cannot make any inferences about possible acceleration of these features between the second and third epochs. 

\clearpage 
%%%%%%%%%%%%%%%%%%%%%%%%%
%Number 23, Canddiate 2, J0136
%%%%%%%%%%%%%%%%%%%%%%%%%%%
\subsubsection{SDSS\,J013656.31$-$004623.8}
\begin{figure*}
\begin{center}
\includegraphics[scale = 0.32, angle = -90, trim = 20 0 0 0, clip]{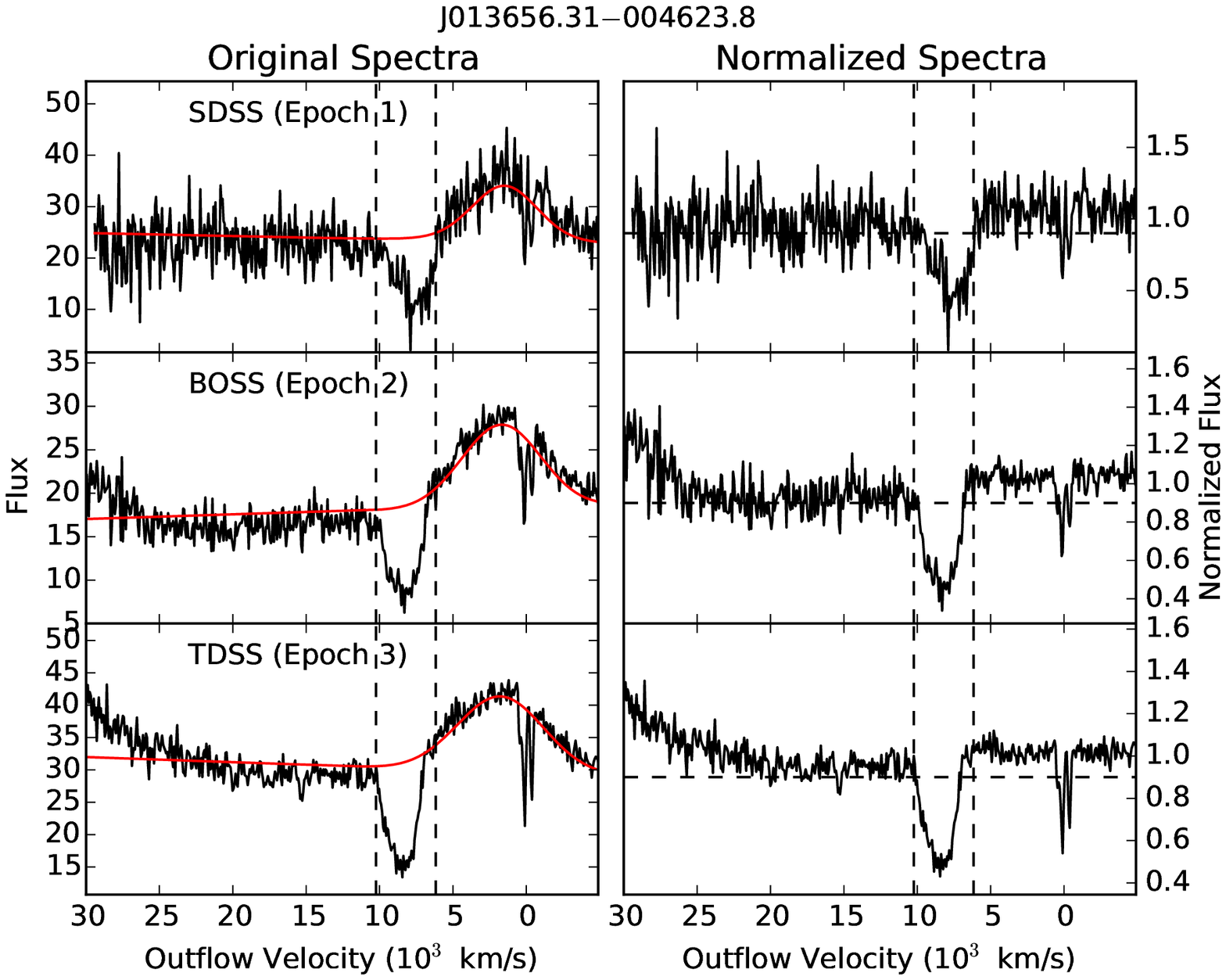} 
\includegraphics[scale = 0.32, angle = -90, trim = 0 0 0 0, clip]{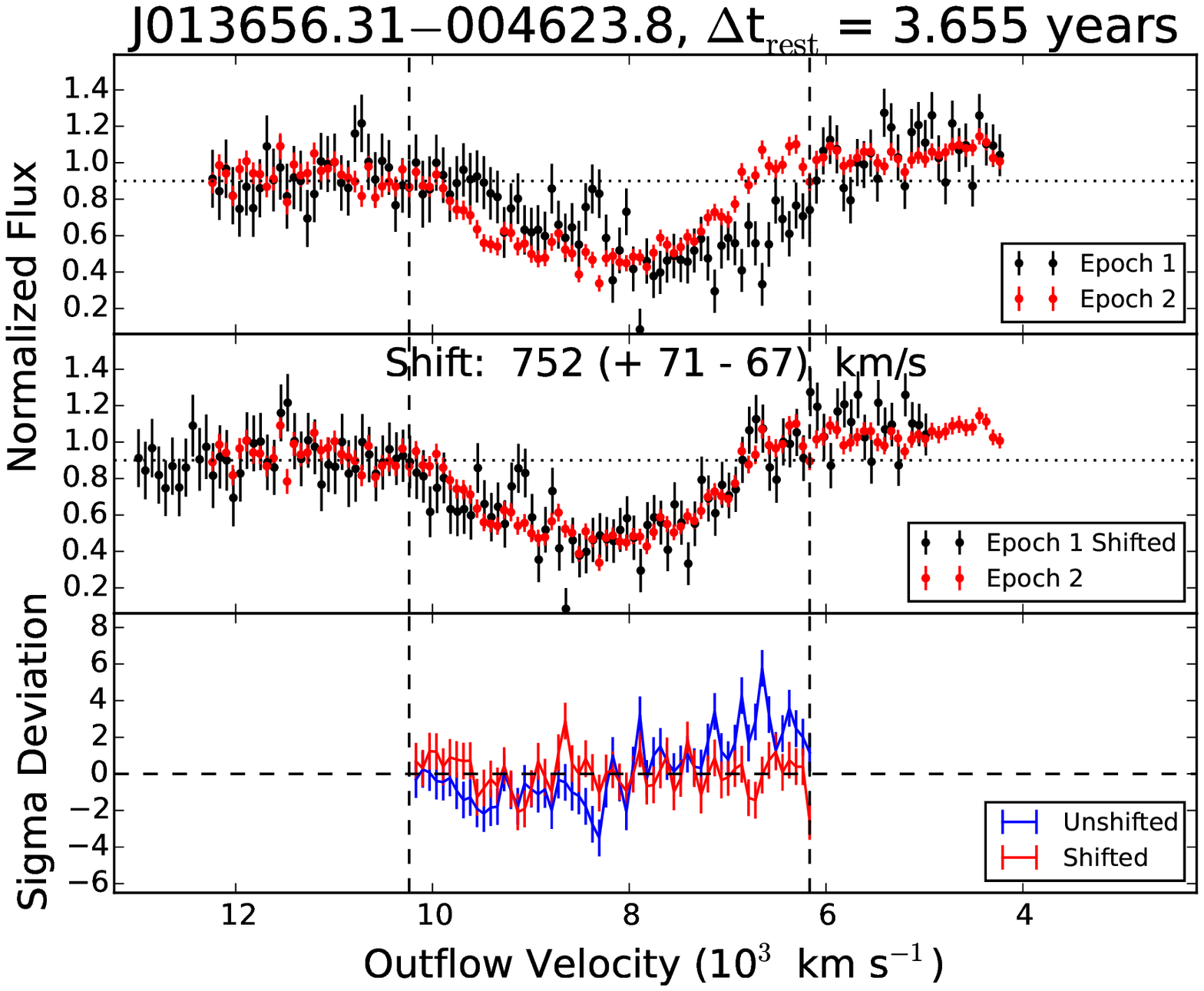} 
\includegraphics[scale = 0.32, angle = -90, trim = 0 0 0 0, clip]{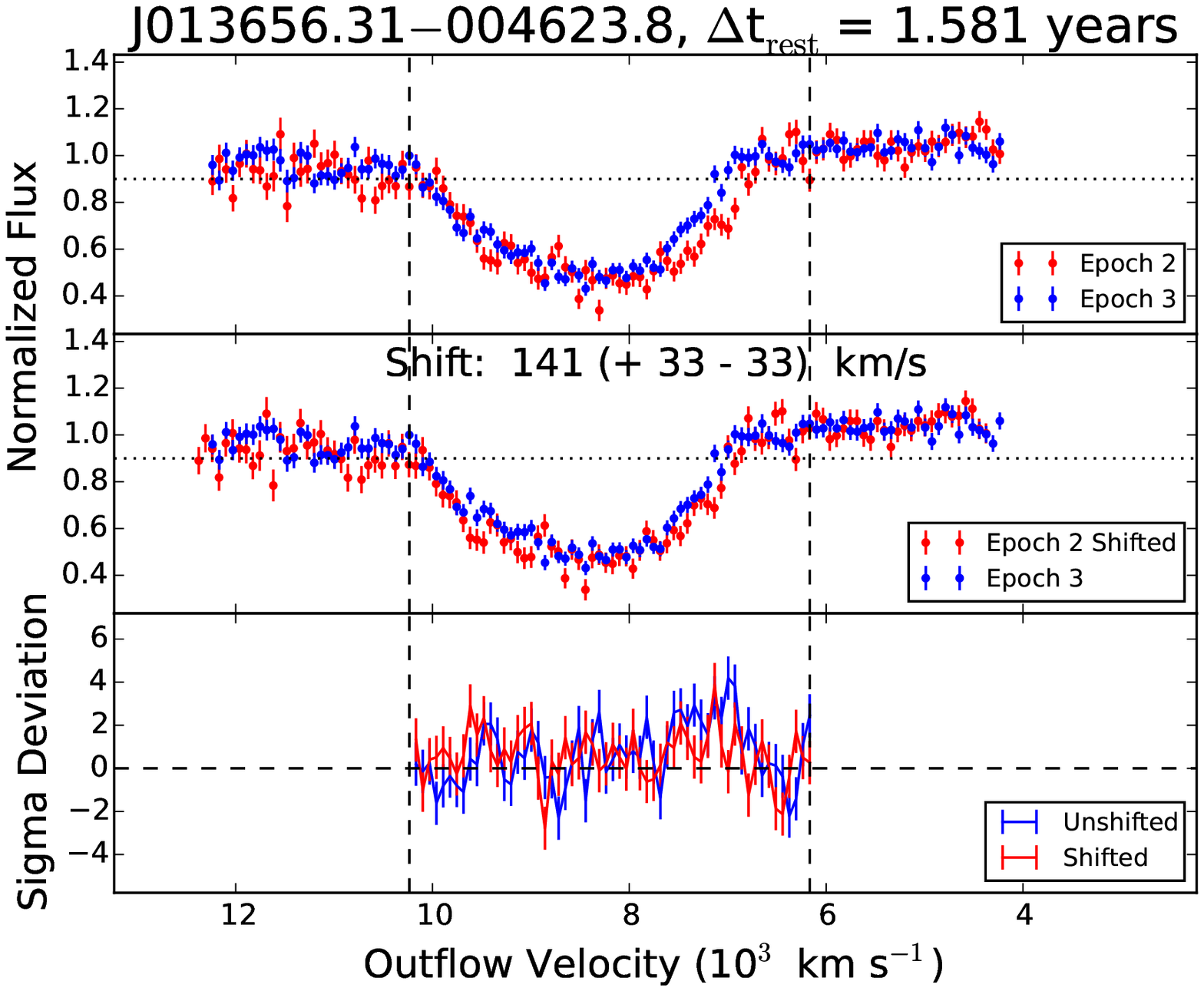} 
\caption{Spectra of the SDSS\,J013656.31$-$004623.8 \civ \ BAL. The left panels show the original spectra, the associated continuum plus \civ \ emission-line fits, and the normalized spectra; vertical dashed lines indicate the velocity range of the BAL in question. The middle and right panels present the pairs of spectra for this target. In all cases, the first epoch (SDSS) is shown in black, the second epoch (BOSS) is shown in red, and the third epoch (TDSS) is shown in blue. The middle subpanel of each panel displays the two spectra with the earlier epoch shifted by the measured velocity shift from the CCF analysis. The bottom subpanel in the middle and right panels shows the deviation (in sigma) between the spectra before and after the shift in red and blue, respectively. The large velocity shifts and systematic trends seen in the deviation between the two profiles that are removed when applying a velocity shift make this BAL our most solid case of possible acceleration. }
\label{fig:accel3}
\end{center}
\end{figure*}

The \civ \ BAL in SDSS\,J013656.31$-$004623.8 (Figure \ref{fig:accel3}) is perhaps our strongest acceleration candidate. We again see a visible shift between Epochs 1 and 2 and Epochs 1 and 3, although the test between Epochs 2 and 3 (right panel of Figure \ref{fig:accel3}) does not formally meet the $\chi^2$ requirements (the $p$-value is too small in the shifted version). This is because the uncertainties are small in the BOSS and TDSS spectra, and there is a very slight narrowing of the BAL profile as well as some isolated, deviant pixels with small uncertainties. However, in this particular system, we see a systematic negative/positive trend in the sigma deviation between all three pairs of spectra (in contrast, the BAL in J012415.53$-$003318.4 only showed this trend between two of the three pairs of spectra), and this trend is removed when the velocity shift is applied in all three cases. We again measure a smaller magnitude of acceleration between Epochs 2 and 3 than between Epochs 1 and 2, indicating that the magnitude of the acceleration is not constant over time. 

\begin{figure}
\begin{center}
\includegraphics[scale = 0.32, angle = 0, trim = 0 0 0 25, clip]{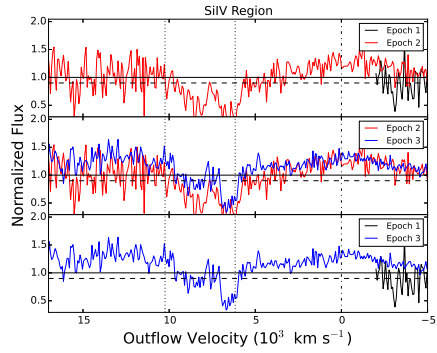} 
\caption{The \siiv \ region of SDSS\,J013656.31$-$004623.8. The top panel shows Epochs 1 and 2, the middle panel shows Epochs 2 and 3, and the bottom panel shows Epochs 1 and 3. The dashed-dotted vertical line indicates the rest wavelength of the \siiv \ line, and the dotted vertical lines represent the velocity range corresponding to $v_{\rm min}$ and $v_{\rm max}$ of the \civ \ BAL for which we detected a velocity shift. Note that we did not fit a line profile to the \siiv \ emission line as we did for the \civ \ line --- hence the raised flux level redward of the absorption complex. Additional flux blueward of the absorption feature (as seen in the bottom panel) is due to the exclusion of the \siiv \ region in the continuum fit, as Epoch 1 did not cover this region (see text for details). }
\label{fig:siiv23}
\end{center}
\end{figure}

J013656.31$-$004623.8 also hosts a \siiv \ BAL complex (Figure \ref{fig:siiv23}) at outflow velocities similar to that of the \civ \ BAL at $v~\sim$~5,000$-$10,000 \kms.  Unfortunately, the \siiv \ region of the spectrum is not covered by the first epoch (hence the lack of comparison spectra over most of the velocity range in the top and bottom panels of Figure \ref{fig:siiv23}). In order to obtain consistent continuum fits between the three epochs in the \civ \ region, we excluded the \siiv \ region in our fits --- thus the continuum fits are less robust in the \siiv \ region.  Additionally, in Epochs 2 and 3 there is decreased S/N in this region due to sky contamination, making it difficult to fit a continuum. Visual inspection suggests that the \siiv \ absorption complex varied in strength during this period; it is unlikely that we would be able to detect any acceleration even if it were present. We also examined the \siiv \ region for Epochs 2 and 3 while including that region, which improved the fit in that region and find that the profile variability persists. We are thus unable to assess any acceleration in this absorption feature regardless of our continuum fits.

We searched for additional BAL features in this spectrum, but the Lyman-$\alpha$/\nv \ region is not covered in any spectrum, and there are no visible absorption features in the \mgii \ region. There is a shallow \aliii \ trough at the same outflow velocities as the \civ \ trough, but it is not deep enough to be considered a bona-fide BAL. This \aliii \ shallow trough shows no visible change between Epochs 1 and 2 and weakens significantly across the entire trough between Epochs 2 and 3, rendering any attempts at measuring acceleration impossible. 

\clearpage
%%%%%%%%%%%%%%%%%%%%%%
%Number 71, Candidate 3, J10914 
%%%%%%%%%%%%%%%%%%%%%%%%
\subsubsection{SDSS\,J091425.73+504854.9} 
\begin{figure*}
\begin{center}
\includegraphics[scale = 0.32, angle = -90, trim = 20 0 0 0, clip]{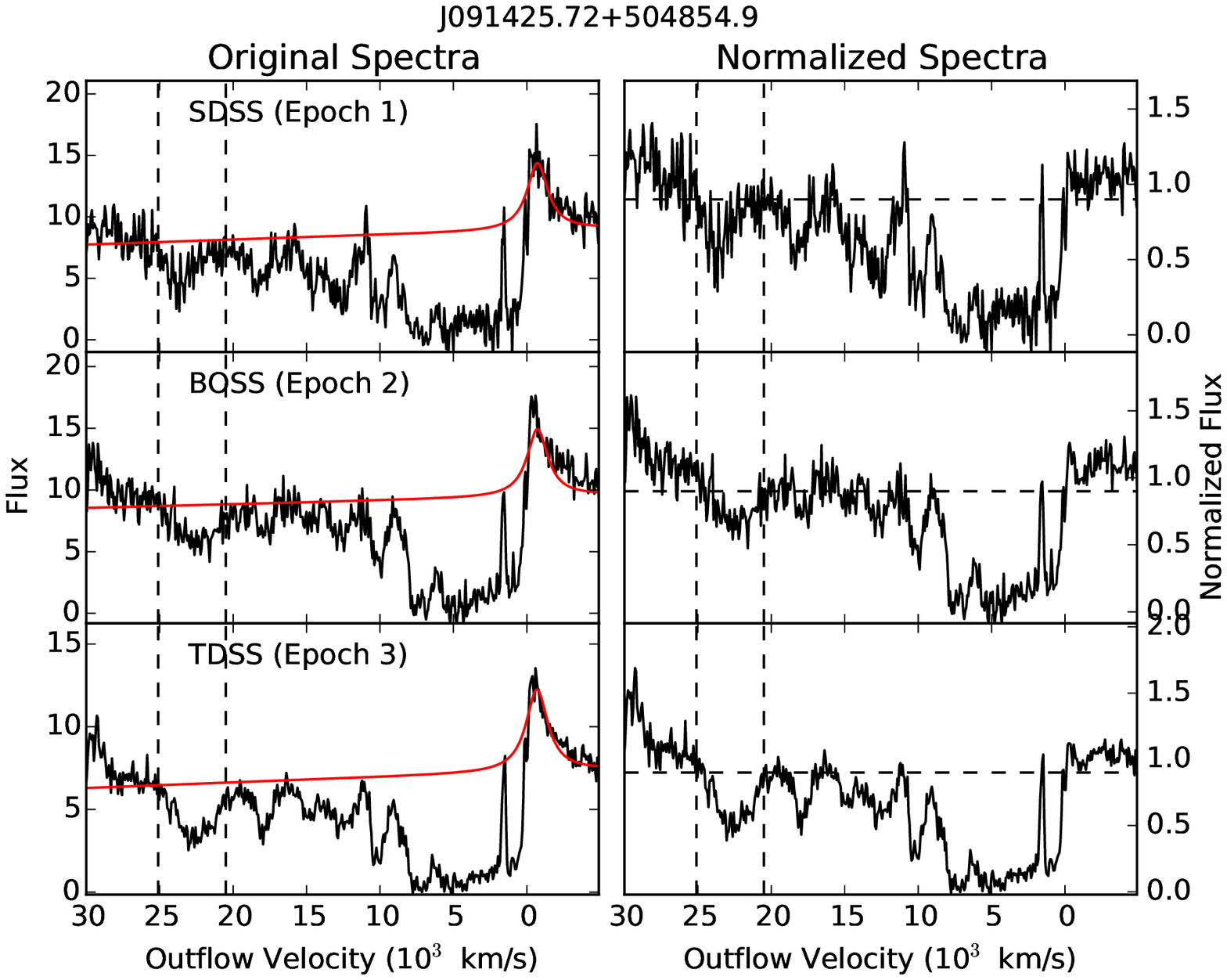} 
\includegraphics[scale = 0.32, angle = -90, trim = 0 0 0 0, clip]{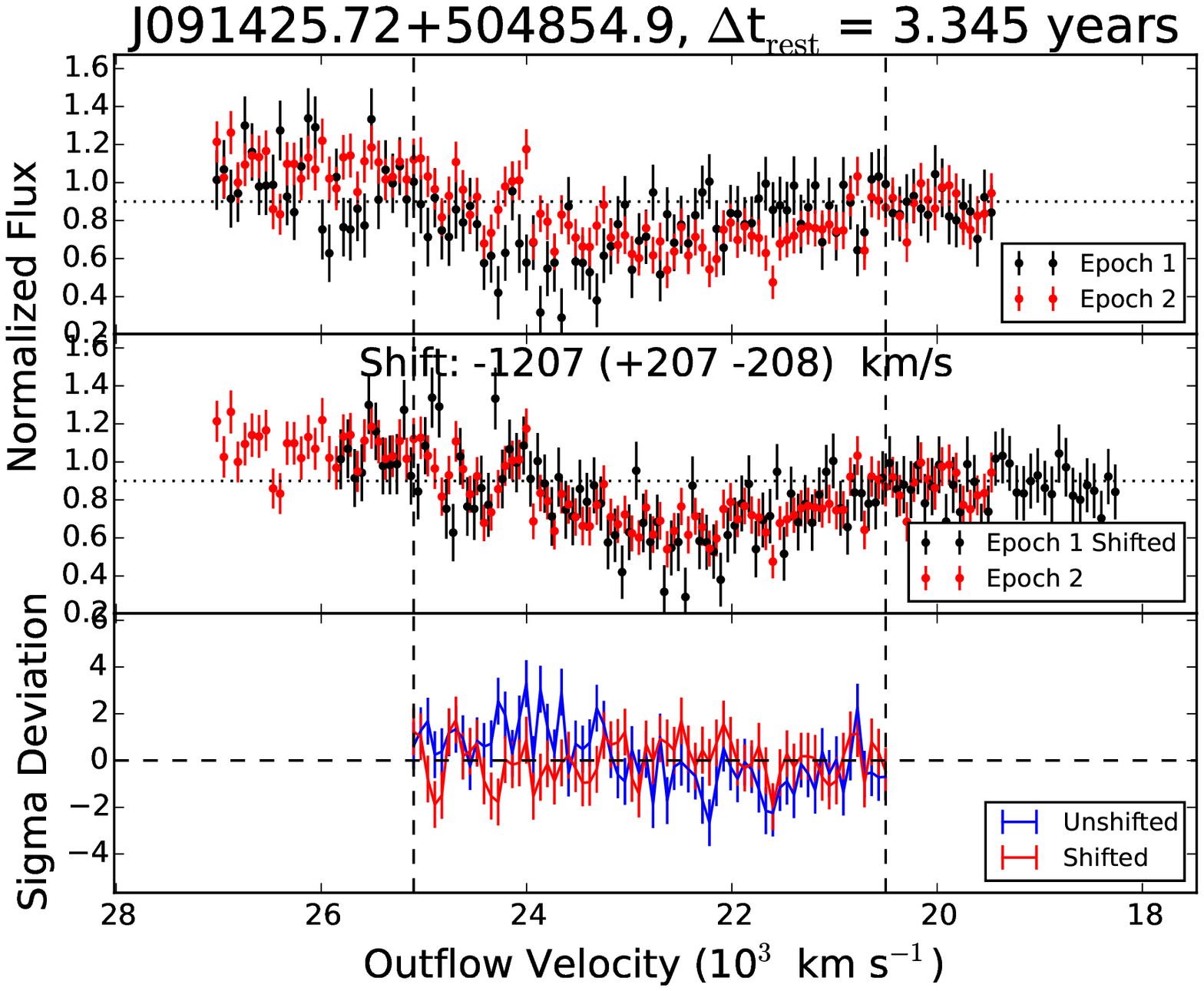} 
\includegraphics[scale = 0.32, angle = -90, trim = 0 0 0 0, clip]{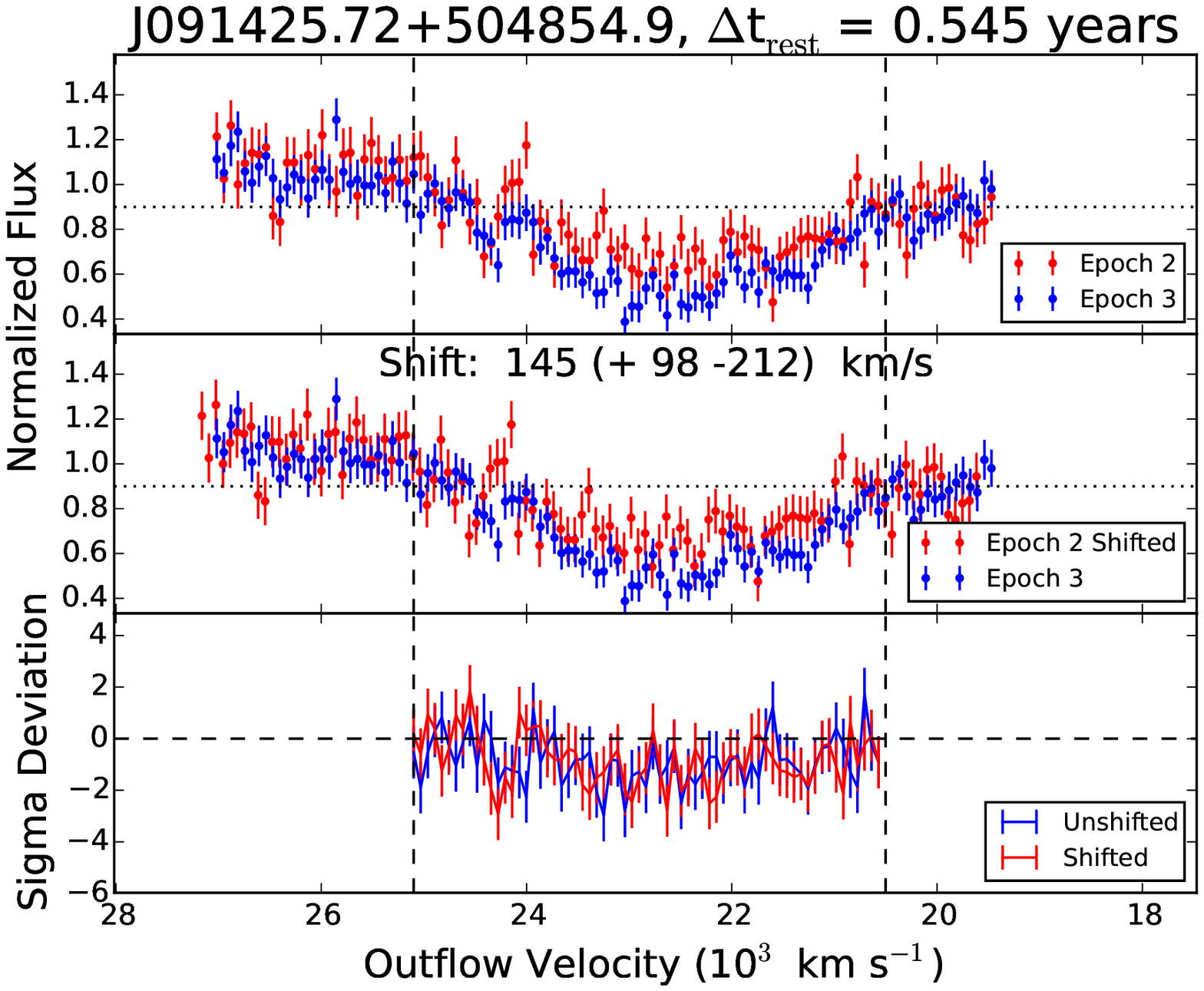} 
\caption{Spectra of the SDSS\,J091425.73+504854.9 \civ \ BAL. The left panels show the original spectra, the associated continuum plus \civ \ emission-line fits, and the normalized spectra; vertical dashed lines indicate the velocity range of the BAL in question. The middle and right panels present the pairs of spectra for this target. In all cases, the first epoch (SDSS) is shown in black, the second epoch (BOSS) is shown in red, and the third epoch (TDSS) is shown in blue. The middle subpanel of each panel displays the two spectra with the earlier epoch shifted by the measured velocity shift from the CCF analysis. The bottom subpanel in the middle and right panels shows the deviation (in sigma) between the spectra before and after the shift in red and blue, respectively. Because the trough weakened between Epochs 2 and 3 and the acceleration appears to have switched from negative to positive, we view this as a less solid case than our first two candidates. }
\label{fig:accel4}
\end{center}
\end{figure*}

SDSS\,J091425.73+504854.9 (Figure \ref{fig:accel4}) is the only case where we detect a shift toward smaller velocities, or deceleration of the BAL. The upper-left panel of Figure \ref{fig:accel4} reveals that this target has several absorption features in the \civ \ line region; the trough in question is the highest-velocity trough, as indicated by the vertical dashed lines. We do not measure significant acceleration in any of the other lower-velocity BAL troughs. 

Due to an increase in trough strength between Epochs 2 and 3 and the relatively short time span between these two epochs, the measurement of acceleration between these two epochs is formally an upper limit; i.e., the acceleration between these two epochs is not detected at a 3$\sigma$ significance. The average acceleration measured between Epochs 1 and 3 is actually smaller in magnitude than the average acceleration measured between Epochs 1 and 2 (which is a longer timescale that includes the first epoch); this result suggests that the change between Epochs 2 and 3 represents positive acceleration rather than a continuing deceleration. The deceleration observed between Epochs 1 and 2 (SDSS and BOSS) formally passes all of our requirements, and we see again that the deviation between these two spectra possesses the same systematic trend we would expect from deceleration; hence, we still consider the first two epochs of this target to be a fairly good indication of deceleration, though admittedly less solid than the acceleration seen in our first two candidates. 

\begin{figure}
\begin{center}
\includegraphics[scale = 0.35, angle = 0, trim = 0 0 0 25, clip]{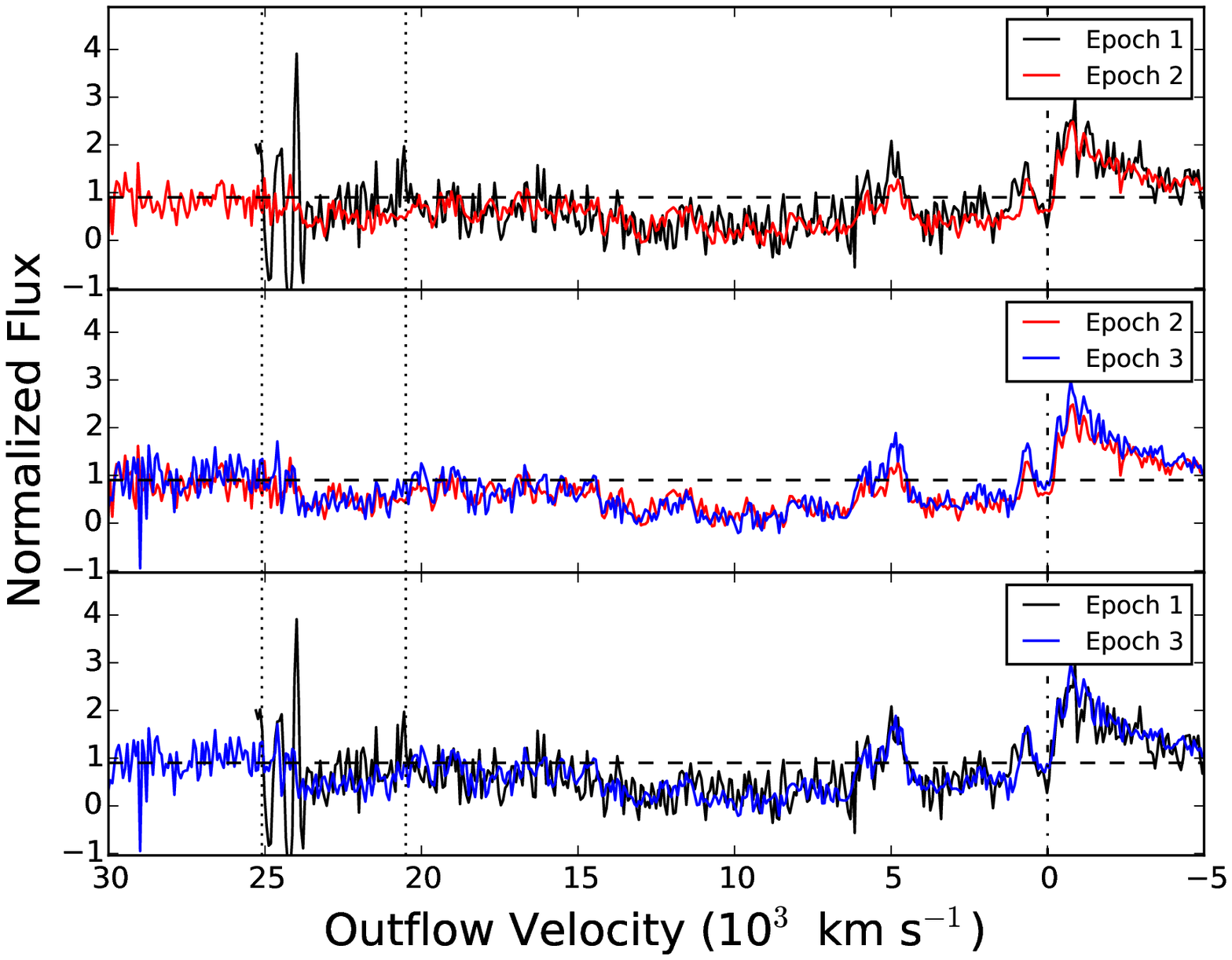} 
\includegraphics[scale = 0.35, angle = 0, trim = 0 0 0 25, clip]{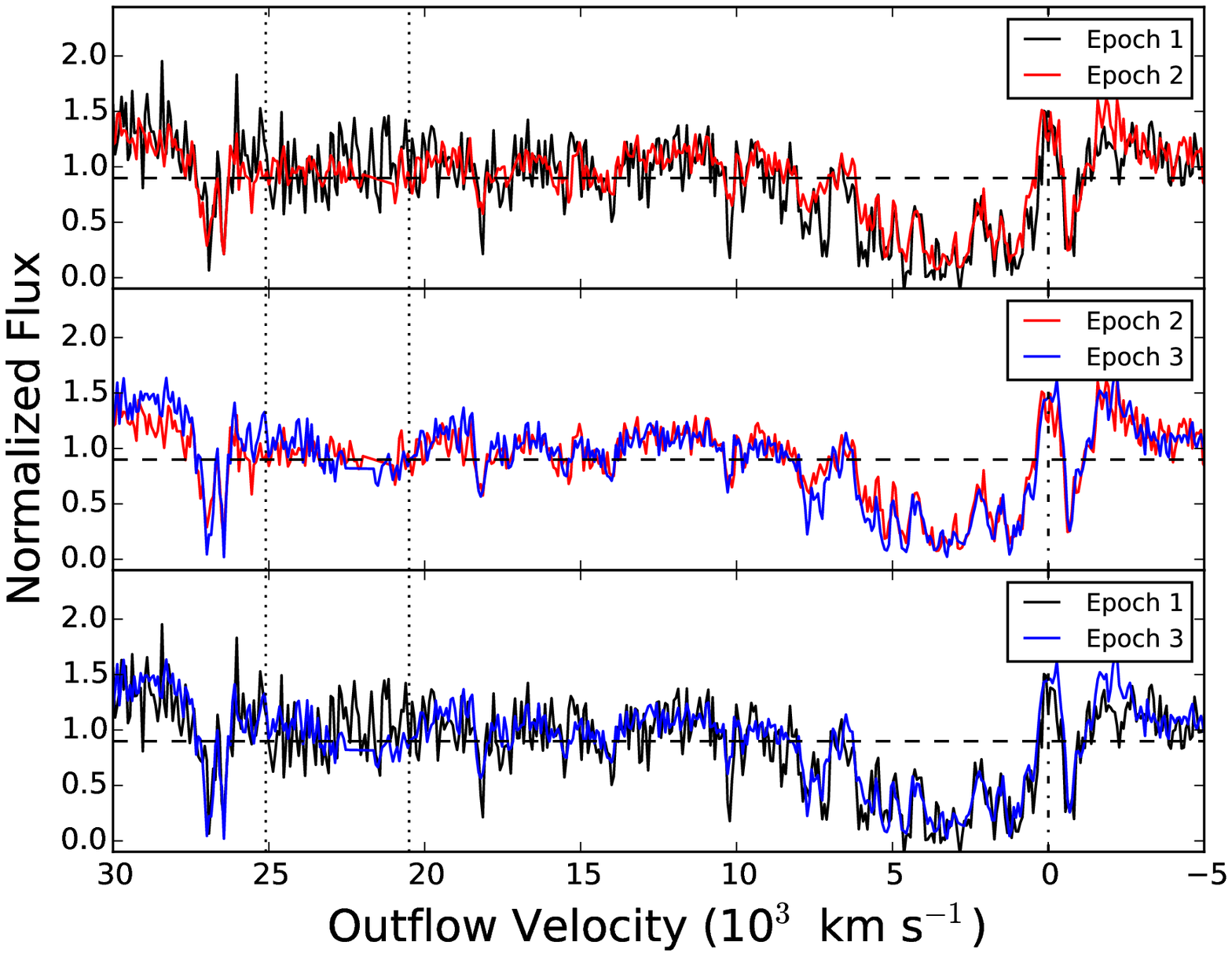} 
\caption{The Lyman-$\alpha$/\nv \ region (left) and \siiv \ region (right) of SDSS\,J1091425.73+504854.9. The top panel shows Epochs 1 and 2, the middle panel shows Epochs 2 and 3, and the bottom panel shows Epochs 1 and 3. The dashed-dotted vertical line indicates the rest wavelength of the \nv \ (left) and \siiv \ (right) lines, and the dotted vertical lines indicate the velocity range corresponding to $v_{\rm min}$ and $v_{\rm max}$ of the \civ \ BAL for which we detected a velocity shift.}
\label{fig:siiv71}
\end{center}
\end{figure}

This target shows absorption in the Lyman-$\alpha$/\nv \ region of the spectrum --- however, velocities corresponding to the \civ \ BAL fall at the very edge of the wavelength coverage in the SDSS spectrum and are subject to significant sky residuals, making exploration of the BAL on long timescales difficult (Figure \ref{fig:siiv71}). There are no \siiv, \aliii, or \mgii \ BALs visible at velocities corresponding to that of the \civ \ BAL identified as an acceleration candidate ($v \sim$ 20,000$-$25,000 \kms), although significant absorption (either BAL or NAL) is visible at low velocities, particularly in the \siiv \ region (right panels of Figure \ref{fig:siiv71}). 

\end{appendix} 
\clearpage 

%%Tables 
\begin{deluxetable*}{lcccccc} 
\tablewidth{0pt} 
\tablecaption{BAL Variability Studies Mentioning Acceleration} 
\tablehead{ 
\colhead{} & 
\colhead{Number of} & 
\colhead{Rest-Frame } & 
\colhead{Redshift } & 
\colhead{Epochs per} & 
\colhead{Spectral} & 
\colhead{Detection}  \\ 
\colhead{Reference} & 
\colhead{Targets} & 
\colhead{Timescales} &  
\colhead{Range} & 
\colhead{Target} & 
\colhead{Resolution}  &
\colhead{Methodology}
} 
\startdata 
This Sample              & 140 &       2.71 -- 5.49 yr                          & 1.6 $< z <$ 4            &   3--4     &       2000     & 1  \\ 
\cite{Gibson08b}      &  13   &  3 -- 6 yr                 &  1.7 $< z <$ 2.8      & 2               &   750 (HET) -- 2000 (SDSS)  & 2  \\ 
\cite{Gibson10}        &  14   &  months -- 6.7 yr    &  $z >$ 2.1                      & 2--4       &   $\sim$1000     & 3  \\ 
\cite{Capellupo12}  &  24   &  15 days -- 8.2 yr  &  1.2 $< z <$ 2.9       & 2--10     &   600 (Lick) -- 2000 (SDSS)      & 4   
%\cite{Lundgren07}    &  29   &   $<$ 0.5 yr                  &   1.7 $< z <$ 3.25   &      2         &   2000     
\enddata 
\tablecomments{Methods used to search for BAL acceleration: 1. Cross correlation and $\chi^2$ test; 2. Examining BALs with steep onsets and assuming an upper limit of 1 \AA \ for velocity shifts; 3. Visual search and shifting by eye; and 4. Unknown.} 
\label{tbl:previous_studies} 
\end{deluxetable*}  

%%%%Table 2, abbreviated    
\begin{table*}   
\caption{Quasar Observation Information}    
\begin{center}   
{\scriptsize    
\begin{tabular}{ccrrcccccc}   
\tableline\tableline \\[-0.3em]   
 & RA   & DEC   &  &  & SDSS &  & BOSS &  \\   
SDSS &  (J2000)  & (J2000)  & &  & Observation & SDSS & Observation & BOSS \\   
Identifier &  (deg) &  (deg) &  $z^a$ & $M_{i}^b$ & Date (UT) & Plate-MJD-Fiber & Date (UT) & Plate-MJD-Fiber  \\   
\\[0.1em]   
\tableline\\[0.1em]   
J001025.91+005447.6  &   2.6080  &  $ 0.9132 $ &  2.860 & $-27.29$  &   2000-09-08  &   0389-51795-0332  &   2010-10-10  &  4218-55479-0592   \\ 
J001130.55+005550.8  &   2.8773  &  $ 0.9308 $ &  2.305 & $-27.43$  &   2000-09-08  &   0389-51795-0339  &   2010-10-09  &  4217-55478-0948   \\ 
J001438.28$-$010750.1  &   3.6595  &  $ -1.1306 $ &  1.805 & $-26.77$  &   2000-09-08  &   0389-51795-0211  &   2010-10-10  &  4218-55479-0218   \\ 
J001502.26+001212.5  &   3.7594  &  $ 0.2035 $ &  2.853 & $-27.39$  &   2000-09-08  &   0389-51795-0465  &   2010-10-10  &  4218-55479-0818   \\ 
J001657.76+005614.6  &   4.2407  &  $ 0.9374 $ &  1.872 & $-26.73$  &   2002-09-01  &   0687-52518-0537  &   2010-10-10  &  4218-55479-0920   \\ 
J001836.40$-$002933.0  &   4.6517  &  $ -0.4925 $ &  1.851 & $-26.55$  &   2000-09-08  &   0389-51795-0066  &   2010-10-10  &  4218-55479-0024   \\ 
J002127.88+010420.2  &   5.3662  &  $ 1.0723 $ &  1.824 & $-26.84$  &   2000-12-22  &   0390-51900-0443  &   2010-10-11  &  4219-55480-0746   \\ 
J002146.71$-$004847.9  &   5.4447  &  $ -0.8133 $ &  2.496 & $-27.15$  &   2000-12-22  &   0390-51900-0180  &   2010-10-11  &  4219-55480-0216   \\ 
J003135.58+003421.2  &   7.8982  &  $ 0.5726 $ &  2.236 & $-27.23$  &   2001-12-19  &   0689-52262-0502  &   2009-12-17  &  3587-55182-0570   \\ 
J003517.95+004333.7  &   8.8248  &  $ 0.7260 $ &  2.917 & $-27.19$  &   2002-09-08  &   1086-52525-0481  &   2009-12-17  &  3587-55182-0722   
\\[0.6em]   
\hline \\[0.6em]   
\end{tabular}   
\begin{tabular}{ccccccc}   
\tableline\tableline \\[-0.3em]   
TDSS & & & & & \\   
Observation & TDSS & SDSS & BOSS & TDSS & Best-fit \\   
Date (UT) & Plate-MJD-Fiber & SNR$_{1700}$ & SNR$_{1700}$ & SNR$_{1700}$ & Profile$^{c}$ \\    
\\[0.1em]   
\tableline\\[0.1em]   
2014-11-27    &   7853-56988-0370   &     6.4  &    20.7  &    16.6  &   G-H  \\ 
2014-11-27    &   7853-56988-0325   &     9.1  &    20.4  &    20.9  &   V  \\ 
2014-11-18    &   7864-56979-0441   &     7.2  &    13.1  &    10.8  &   V  \\ 
2014-11-18    &   7864-56979-0618   &    10.6  &    21.4  &    24.4  &   2-G  \\ 
2014-11-18    &   7864-56979-0710   &     7.0  &    14.7  &    17.9  &   2-G  \\ 
2014-11-18    &   7864-56979-0220   &     7.2  &    22.0  &    18.6  &   2-G  \\ 
2014-11-28    &   7854-56989-0326   &    18.6  &    26.5  &    23.1  &   G-H  \\ 
2014-10-25    &   7865-56955-0748   &    15.0  &    18.7  &    21.3  &   V  \\ 
2014-12-15    &   7868-57006-0512   &    15.0  &    24.5  &    27.6  &   V  \\ 
2014-12-20    &   7855-57011-0082   &    12.3  &    14.6  &     9.5  &   V  
\\[0.6em]
\hline    
\end{tabular}}   
\end{center}   
\tablecomments{This table is available in its entirety in a machine-readable form in the online journal. A portion is shown here for guidance regarding its form and content. (a) Redshifts are from \cite{Hewett10}. (b) Absolute magnitudes are from the DR7 quasar catalog (\citealt{Schneider10}). (c) Best-fit profiles for the \civ \ emission line (see Section \ref{sec:confits}): G-H~=~Gauss-Hermite, 2-G~=~Double-Gaussian, V~=~Voigt profile, and N~=~No line present. } 
\label{tbl:sample_info}     
\end{table*}   

\clearpage 

%%%%Candidate info table (Table 3)

\setlength{\tabcolsep}{3pt} 
\begin{table*} 
\caption{BAL-Acceleration Candidate Measurements}  
\begin{center} 
{\scriptsize  
\begin{tabular}{cccccccrccccc}  
\tableline\tableline \\[-0.3em] 
SDSS &  First  & Second  & $\Delta t^{a}$ & $v_{\rm min}$ & $v_{\rm max}$ & EW$^{b}$ & Vel Shift$^{c}$ &  Accel$^{c}$  &  \multicolumn{2}{c}{Unshifted} & \multicolumn{2}{c}{Shifted} \\  
\cmidrule{10-11}\cmidrule{12-13}  
ID &  Epoch & Epoch & (years) & (km s$^{-1}$)  & (km s$^{-1}$) & (\AA) & (km s$^{-1}$)  & (cm s$^{-2}$) & $\chi^2$/DOF & $p$ & $\chi^2$/DOF & $p$  \\ 
\\[0.05em] 
\tableline\\[0.05em] 
J012415.53$-$003318.4   &  SDSS   &  BOSS &   2.424   &        16531  &      20519  &  3.20  &  $        691^{+  159}_{-  245}$  & $0.904^{+0.207}_{-0.321}$ &  1.61  &  0.00  &  0.80   &   0.86 \\ 
\nodata  &  BOSS   &  TDSS &   1.252   &   \nodata  & \nodata  & \nodata & $        145^{+   69}_{-   38}$  & $0.368^{+0.174}_{-0.097}$ &  3.06  &  0.00  &  2.71   &   0.00 \\ 
\nodata  &   SDSS   &  TDSS &   3.677   &  \nodata  & \nodata  & \nodata & $        731^{+  166}_{-  150}$  & $0.630^{+0.143}_{-0.129}$ &  1.92  &  0.00  &  0.69   &   0.96\\ 
J013656.31$-$004623.8   &  SDSS   &  BOSS &   3.655   &         6166  &      10236  &  5.85  &  $        752^{+   71}_{-   67}$  & $0.652^{+0.061}_{-0.058}$ &  3.43  &  0.00  &  0.98   &   0.51 \\ 
\nodata  &  BOSS   &  TDSS &   1.581   &   \nodata  & \nodata  & \nodata & $        141^{+   33}_{-   33}$  & $0.282^{+0.067}_{-0.067}$ &  2.64  &  0.00  &  1.72   &   0.00 \\ 
\nodata  &   SDSS   &  TDSS &   5.236   &  \nodata  & \nodata  & \nodata & $        894^{+   70}_{-   70}$  & $0.541^{+0.043}_{-0.042}$ &  4.37  &  0.00  &  1.12   &   0.25\\ 
J091425.72+504854.9   &  SDSS   &  BOSS &   3.345   &        20501  &      25100  &  4.01 &  $      -1207^{+  207}_{-  208}$  & $-1.144^{+0.196}_{-0.197}$ &  1.66  &  0.00  &  0.81   &   0.86 \\ 
\nodata  &  BOSS   &  TDSS &   0.545   &   \nodata  & \nodata  & \nodata & $        145^{+   98}_{-  212}$  & $0.844^{+0.567}_{-1.233}$ &  1.61  &  0.00  &  1.68   &   0.00 \\ 
\nodata  &   SDSS   &  TDSS &   3.891   &   \nodata  & \nodata  &  \nodata  & $      -1012^{+  227}_{-  296}$  &  $-0.825^{+0.185}_{-0.241}$ &  2.06  &  0.00  &  1.23   &   0.10 
\\[0.6em]   
\hline    
\end{tabular}}   
\label{tbl:candidates}     
\tablecomments{(a) $\Delta t$ is measured in the quasar rest frame. (b) The quoted equivalent width (EW) is the average between all three epochs.   
(c) Positive values refer to acceleration, or a shift toward higher velocities; negative values refer to deceleration, or a shift toward smaller velocities. }  
\end{center}   
\end{table*}     
\setlength{\tabcolsep}{6pt} 

\clearpage 

%%%%Upper limits table (table 4) 
\setlength{\tabcolsep}{5pt} 
\renewcommand{\thefootnote}{\alph{footnote}} 
\scriptsize  
\begin{center} 
\begin{longtable}{lccrrrrrrcc}  
\caption{BAL Information and Acceleration Upper Limits} 
\\[0.5ex]\hline 
\multicolumn{1}{c}{SDSS} &  & $\Delta t^{b}$ & $v_{\rm min}$$^{c}$ & $v_{\rm max}$$^{c}$ & EW 1$^{d}$ & EW 2$^{d}$ & 
EW 3$^{d}$ & Velocity & Accel & Decel \\ 
\multicolumn{1}{c}{ID} &  $z$$^{a}$ &  (years) &  (km s$^{-1}$) & (km s$^{-1}$) & (\AA) & (\AA) & (\AA) & 
Shift$^{e,f}$ & (cm s$^{-2}$)$^{f, g}$ & (cm s$^{-2}$)$^{f, g}$ 
\\[0.05em] 
\hline\\[0.05em] 
\endfirsthead 
\\[0.5ex]\hline 
\multicolumn{1}{c}{SDSS} &  & $\Delta t^{b}$ & $v_{\rm min}$$^{c}$ & $v_{\rm max}$$^{c}$ & EW 1$^{d}$ & EW 2$^{d}$ & 
EW 3$^{d}$ & Velocity & Accel & Decel \\ 
\multicolumn{1}{c}{ID} &  $z$$^{a}$ &  (years) &  (km s$^{-1}$) & (km s$^{-1}$) & (\AA) & (\AA) & (\AA) & 
Shift$^{e,f}$ & (cm s$^{-2}$)$^{f, g}$ & (cm s$^{-2}$)$^{f, g}$ 
\\[0.05em] 
\hline\\[0.05em] 
\endhead 
\\[0.05em] 
\hline\\[0.05em] 
\endfoot 
\\[0.6em]   
\hline    
\endlastfoot 
\\[0.05em] 
\multicolumn{11}{c}{Full-Trough Upper Limits} \\ 
\\[0.05em] 
\tableline\\[0.05em] 
J001130.55+005550.8   &  2.305   &   4.305   &           508 &       5478  &         8.7   &        8.1  &        8.2 &  $       -34^{+   67}_{-   66}$  &  $<$ 0.049  &  $<$ 0.048\\ 
J001502.26+001212.5   &  2.853   &   3.687   &          5228 &       9919  &         7.7   &        6.2  &        7.6 &  $         1^{+  135}_{-  135}$  &  $<$ 0.116  &  $<$ 0.116\\ 
J003517.95+004333.7   &  2.917   &   3.138   &         16529 &      19624  &         2.2   &        2.2  &        1.5 &  $       618^{+  831}_{-  963}$  &  $<$ 0.839  &  $<$ 0.974\\ 
J003551.98+005726.4   &  1.906   &   4.915   &          4010 &       8977  &        12.6   &        9.4  &       11.5 &  $       -34^{+  131}_{-  129}$  &  $<$ 0.084  &  $<$ 0.083\\ 
J004732.73+002111.3   &  2.873   &   3.694   &          7108 &      10626  &         4.2   &        0.0  &        3.4 &  $        66^{+  103}_{-  163}$  &  $<$ 0.089  &  $<$ 0.140\\ 
J012633.53+003509.9   &  2.137   &   4.518   &          1295 &       3849  &         7.7   &        6.9  &        7.3 &  $       -34^{+   34}_{-   65}$  &  $<$ 0.024  &  $<$ 0.046\\ 
J023252.80$-$001351.1   &  2.028   &   4.722   &          3263 &       5817  &         7.5   &        6.2  &        7.1 &  $         2^{+   65}_{-   37}$  &  $<$ 0.043  &  $<$ 0.025\\ 
J024230.65$-$000029.7   &  2.495   &   3.813   &          5724 &      19571  &        18.0   &       22.9  &       21.6 &  $       -32^{+  334}_{-  269}$  &  $<$ 0.278  &  $<$ 0.223\\ 
J025331.92+001624.7   &  1.825   &   5.012   &          3275 &       6588  &         4.7   &        6.8  &        4.2 &  $        64^{+  165}_{-  202}$  &  $<$ 0.105  &  $<$ 0.128\\ 
J080944.76+483451.9   &  3.247   &   3.295   &         16269 &      22524  &         5.8   &        5.7  &        5.1 &  $        33^{+  102}_{-  102}$  &  $<$ 0.098  &  $<$ 0.098\\ 
J090353.93+565346.5   &  2.053   &   4.592   &           293 &       2915  &         7.5   &        7.4  &        7.5 &  $       -37^{+   69}_{-   99}$  &  $<$ 0.048  &  $<$ 0.068\\ 
J090931.86+541140.8   &  1.718   &   5.226   &          1277 &       5073  &        12.1   &       11.4  &       10.9 &  $        32^{+  135}_{-  125}$  &  $<$ 0.082  &  $<$ 0.076\\ 
J091425.72+504854.9   &  2.341   &   3.891   &         17063 &      19401  &         2.6   &        0.0  &        2.3 &  $      -311^{+  500}_{-  724}$  &  $<$ 0.408  &  $<$ 0.590\\ 
 J091425.72+504854.9  &  \nodata  &  \nodata    &          1619  &      10795  &        27.1  &        21.2  &       25.9 &  $        62^{+   37}_{-   93}$ &  $<$ 0.030  &  $<$ 0.076 \\ 
J094017.63+445431.4   &  1.718   &   4.411   &          1598 &       5463  &        10.7   &       10.2  &       10.4 &  $      -102^{+  199}_{-  164}$  &  $<$ 0.143  &  $<$ 0.118\\ 
J101547.39+512615.8   &  2.764   &   3.119   &         12962 &      16267  &         1.7   &        3.5  &        1.9 &  $       828^{+ 1104}_{- 2622}$  &  $<$ 1.122  &  $<$ 2.666\\ 
 J101547.39+512615.8  &  \nodata  &  \nodata    &         10136  &      12479  &         3.1  &         3.3  &        2.5 &  $        35^{+  164}_{-  173}$ &  $<$ 0.167  &  $<$ 0.176 \\ 
J105416.51+512326.0   &  2.341   &   3.546   &          1943 &       5256  &         4.9   &        4.8  &        4.4 &  $         0^{+   32}_{-    3}$  &  $<$ 0.029  &  $<$ 0.002\\ 
J112554.71+571841.4   &  2.956   &   2.838   &         19081 &      22790  &         3.8   &        3.0  &        3.4 &  $       -32^{+   65}_{-  198}$  &  $<$ 0.073  &  $<$ 0.221\\ 
 J112554.71+571841.4  &  \nodata  &  \nodata    &          6887  &      15434  &        14.5  &        13.9  &       14.1 &  $       -36^{+  102}_{-  101}$ &  $<$ 0.113  &  $<$ 0.113 \\ 
J112736.70+485939.3   &  1.845   &   4.320   &         11510 &      27248  &        34.2   &       35.5  &       42.2 &  $       -67^{+ 1034}_{- 1135}$  &  $<$ 0.759  &  $<$ 0.833\\ 
 J112736.70+485939.3  &  \nodata  &  \nodata    &           540  &       3645  &        10.1  &         8.6  &       10.4 &  $        -0^{+  131}_{-  102}$ &  $<$ 0.096  &  $<$ 0.075 \\ 
J113009.40+495247.9   &  2.087   &   3.986   &          3526 &       9391  &        17.1   &       16.2  &       16.5 &  $       -33^{+   98}_{-   69}$  &  $<$ 0.078  &  $<$ 0.055\\ 
J114013.07+515944.9   &  2.880   &   3.306   &         12017 &      16081  &         1.9   &        3.2  &        2.8 &  $      -552^{+ 1863}_{- 1380}$  &  $<$ 1.787  &  $<$ 1.323\\ 
 J114013.07+515944.9  &  \nodata  &  \nodata    &          6431  &       8570  &         3.3  &         3.3  &        3.8 &  $         6^{+  157}_{-   73}$ &  $<$ 0.151  &  $<$ 0.070 \\ 
J120139.34+491327.9   &  2.882   &   3.312   &         11096 &      22861  &        18.8   &       18.4  &       19.7 &  $        71^{+  135}_{-  135}$  &  $<$ 0.129  &  $<$ 0.129\\ 
J123224.36+481626.4   &  3.116   &   2.713   &           870 &       3562  &         3.2   &        3.5  &        3.5 &  $       -37^{+  104}_{-  163}$  &  $<$ 0.122  &  $<$ 0.190\\ 
\tableline\\[0.01em] 
\\[0.05em] 
\multicolumn{11}{c}{Partial-Trough Upper Limits} \\ 
\\[0.05em] 
\tableline\\[0.01em] 
J001025.91+005447.6   &  2.860   &   3.686   &   $       7369 $&       8202  &         9.3   &       12.1  &       13.0 &  $        35^{+  862}_{- 1001}$  &  $<$ 0.741  &  $<$ 0.862\\ 
J001438.28$-$010750.1   &  1.805   &   5.063   &   $       -132 $&       1048  &         9.1   &       20.4  &       13.2 &  $         2^{+   33}_{-   34}$  &  $<$ 0.021  &  $<$ 0.021\\ 
J003135.58+003421.2   &  2.236   &   4.016   &   $       4272 $&       6943  &        21.4   &       21.5  &       17.5 &  $        35^{+   67}_{-   66}$  &  $<$ 0.053  &  $<$ 0.052\\ 
J011227.60$-$011221.6   &  1.759   &   5.150   &   $       5324 $&       8394  &        14.6   &       16.9  &       14.0 &  $        69^{+   70}_{-  100}$  &  $<$ 0.043  &  $<$ 0.061\\ 
J011948.52+004355.9   &  1.753   &   5.187   &   $       -848 $&        680  &         4.1   &       15.0  &        8.2 &  $        32^{+   33}_{-   63}$  &  $<$ 0.020  &  $<$ 0.038\\ 
J012913.71+011428.0   &  1.779   &   5.116   &   $        265 $&       1528  &        17.5   &       12.6  &       13.4 &  $       -34^{+  105}_{-  102}$  &  $<$ 0.065  &  $<$ 0.063\\ 
J014548.55$-$000812.5   &  2.804   &   2.853   &   $      16965 $&      20051  &         7.0   &        7.3  &        8.0 &  $       -73^{+  137}_{-  131}$  &  $<$ 0.152  &  $<$ 0.145\\ 
J015024.44+004433.0   &  2.001   &   4.662   &   $        341 $&       4695  &        24.7   &       25.6  &       21.7 &  $        30^{+  170}_{-  160}$  &  $<$ 0.116  &  $<$ 0.109\\ 
J020006.31$-$003709.7   &  2.142   &   4.461   &   $       6906 $&      11208  &        42.6   &       43.9  &       38.6 &  $         1^{+   49}_{-   35}$  &  $<$ 0.035  &  $<$ 0.025\\ 
J021606.40+011509.5   &  2.231   &   4.432   &   $       3088 $&       3639  &         5.3   &        0.0  &        4.5 &  $       -32^{+  998}_{-  865}$  &  $<$ 0.714  &  $<$ 0.619\\ 
J022844.09+000217.0   &  2.726   &   3.802   &   $       6731 $&       8382  &         9.4   &       11.4  &       10.5 &  $         1^{+  204}_{-  129}$  &  $<$ 0.170  &  $<$ 0.107\\ 
J023647.11$-$003124.2   &  2.400   &   3.254   &   $       2567 $&       4153  &        10.1   &        9.4  &        8.7 &  $         2^{+   68}_{-   68}$  &  $<$ 0.066  &  $<$ 0.066\\ 
J024304.68+000005.4   &  2.007   &   4.756   &   $       4128 $&       5389  &         7.4   &        4.8  &        8.2 &  $       -35^{+   97}_{-   67}$  &  $<$ 0.065  &  $<$ 0.045\\ 
J024701.19+000330.2   &  2.149   &   4.497   &   $        471 $&       1822  &         5.3   &        3.6  &        4.6 &  $       131^{+  105}_{-  164}$  &  $<$ 0.074  &  $<$ 0.116\\ 
J025042.45+003536.7   &  2.393   &   4.219   &   $       1278 $&       3895  &        36.5   &       29.2  &       28.3 &  $        31^{+    4}_{-   62}$  &  $<$ 0.003  &  $<$ 0.047\\ 
J025812.86+010603.3   &  2.222   &   4.090   &   $       1457 $&       3251  &        21.9   &       22.2  &       17.4 &  $        -4^{+   69}_{-   98}$  &  $<$ 0.054  &  $<$ 0.076\\ 
J080006.33+443555.6   &  2.519   &   4.041   &   $        -14 $&       2663  &        20.8   &       19.8  &       21.1 &  $         0^{+   66}_{-   66}$  &  $<$ 0.052  &  $<$ 0.052\\ 
J082804.54+445256.8   &  2.079   &   4.518   &   $       1106 $&       2702  &         7.3   &        7.1  &        6.9 &  $        -1^{+   33}_{-   34}$  &  $<$ 0.023  &  $<$ 0.024\\ 
J084957.68+543529.7   &  3.861   &   2.910   &   $      13359 $&      19220  &        12.5   &       10.0  &        9.8 &  $        -0^{+   68}_{-   68}$  &  $<$ 0.075  &  $<$ 0.075\\ 
J090731.56+581142.0   &  1.894   &   4.882   &   $       2632 $&       3683  &        10.4   &        0.0  &        3.6 &  $       -68^{+  234}_{-  239}$  &  $<$ 0.152  &  $<$ 0.155\\ 
J090814.47+550700.4   &  1.931   &   4.797   &   $       -556 $&       1127  &        15.4   &       15.6  &       17.2 &  $        -2^{+  100}_{-  101}$  &  $<$ 0.066  &  $<$ 0.067\\ 
J091035.45+574643.3   &  2.228   &   4.358   &   $       5326 $&       6830  &         5.5   &        4.6  &        5.6 &  $       -31^{+   32}_{-    5}$  &  $<$ 0.023  &  $<$ 0.003\\ 
J091035.45+574643.3   &  2.228   &   4.358   &   $       1737 $&       3069  &        10.1   &        9.3  &        9.8 &  $        -2^{+   33}_{-   63}$  &  $<$ 0.024  &  $<$ 0.046\\ 
J091307.83+442014.3   &  2.946   &   3.299   &   $       9480 $&      12486  &        15.0   &       15.8  &       16.2 &  $       -34^{+  123}_{-  164}$  &  $<$ 0.118  &  $<$ 0.158\\ 
J091307.83+442014.3   &  2.946   &   3.299   &   $       3125 $&       5265  &         6.2   &        6.0  &        6.2 &  $        -2^{+    2}_{-    2}$  &  $<$ 0.002  &  $<$ 0.002\\ 
J091425.72+504854.9   &  2.341   &   3.891   &   $       1620 $&       3635  &        27.1   &       21.2  &       25.9 &  $         4^{+   98}_{-   66}$  &  $<$ 0.080  &  $<$ 0.054\\ 
J092536.61+540227.8   &  2.031   &   4.271   &   $      18764 $&      19949  &        11.2   &       10.3  &       12.4 &  $       183^{+  266}_{-  293}$  &  $<$ 0.197  &  $<$ 0.218\\ 
J093819.08+503912.0   &  1.676   &   4.196   &   $        619 $&       1656  &         6.1   &        9.4  &        9.3 &  $       -37^{+  101}_{-   99}$  &  $<$ 0.076  &  $<$ 0.075\\ 
J095422.68+524903.8   &  2.333   &   3.776   &   $        895 $&       4519  &        22.2   &       27.0  &       25.8 &  $        64^{+  130}_{-   79}$  &  $<$ 0.109  &  $<$ 0.066\\ 
J095712.63+512058.9   &  2.115   &   4.032   &   $       2488 $&       5012  &        32.7   &       28.6  &       26.5 &  $        61^{+  134}_{-  160}$  &  $<$ 0.105  &  $<$ 0.125\\ 
J101108.88+515553.7   &  2.465   &   3.668   &   $        783 $&       3891  &        18.5   &       16.3  &       18.7 &  $        -1^{+  133}_{-  131}$  &  $<$ 0.115  &  $<$ 0.114\\ 
J102850.30+511053.1   &  2.426   &   3.477   &   $      14463 $&      17547  &        18.8   &       12.5  &       15.6 &  $       -34^{+  102}_{-  100}$  &  $<$ 0.093  &  $<$ 0.091\\ 
J102850.30+511053.1   &  2.426   &   3.477   &   $       2531 $&       4233  &        16.4   &       11.9  &       12.3 &  $        -1^{+   34}_{-   64}$  &  $<$ 0.031  &  $<$ 0.058\\ 
J104645.83+512333.3   &  1.735   &   4.409   &   $       8605 $&       9671  &         8.2   &        9.6  &        9.4 &  $        31^{+  616}_{-  443}$  &  $<$ 0.443  &  $<$ 0.318\\ 
J110920.88+545234.2   &  1.946   &   4.172   &   $       1101 $&       3190  &         7.0   &        6.6  &        6.6 &  $        -1^{+    2}_{-   32}$  &  $<$ 0.002  &  $<$ 0.024\\ 
J112440.80+550233.6   &  2.933   &   3.082   &   $       1068 $&       2279  &         5.8   &        5.8  &        5.8 &  $      -131^{+  230}_{-  107}$  &  $<$ 0.237  &  $<$ 0.110\\ 
J113009.40+495247.9   &  2.087   &   3.986   &   $      10963 $&      14856  &        20.3   &       28.3  &       27.5 &  $        62^{+  110}_{-  165}$  &  $<$ 0.087  &  $<$ 0.131\\ 
J113120.04+505615.0   &  2.002   &   4.353   &   $       6884 $&       9374  &        27.0   &       13.9  &       11.9 &  $       161^{+  144}_{-  164}$  &  $<$ 0.105  &  $<$ 0.120\\ 
J113152.57+584510.2   &  2.262   &   3.966   &   $       1234 $&       3351  &        10.1   &        9.7  &       10.0 &  $         2^{+  227}_{-  196}$  &  $<$ 0.182  &  $<$ 0.156\\ 
J113406.88+525958.9   &  1.765   &   4.640   &   $       3970 $&       5026  &        10.8   &        8.6  &        7.7 &  $       -68^{+  168}_{-  105}$  &  $<$ 0.115  &  $<$ 0.072\\ 
J114354.81+541623.1   &  2.858   &   3.064   &   $       4601 $&       6541  &        17.3   &       15.5  &       13.1 &  $        67^{+  202}_{-  266}$  &  $<$ 0.209  &  $<$ 0.276\\ 
J131504.49+500239.5   &  3.283   &   2.823   &   $       -247 $&       2666  &         9.5   &        9.6  &        9.7 &  $        -0^{+   34}_{-   34}$  &  $<$ 0.038  &  $<$ 0.038\\ 
J132815.32+490428.4   &  2.189   &   3.783   &   $       -242 $&       2179  &        43.1   &       30.5  &       28.1 &  $        -0^{+   97}_{-   65}$  &  $<$ 0.081  &  $<$ 0.054\\ 
J134458.82+483457.5   &  2.048   &   3.928   &   $       2837 $&       5836  &        25.4   &       41.7  &       39.5 &  $        35^{+   68}_{-   68}$  &  $<$ 0.055  &  $<$ 0.055\\ 
J135123.47+474712.1   &  1.948   &   4.062   &   $        479 $&       2380  &        11.1   &       12.9  &       12.4 &  $        -0^{+   34}_{-   65}$  &  $<$ 0.027  &  $<$ 0.050\\ 
J141421.53+522940.0   &  2.041   &   4.028   &   $       -588 $&       3109  &        30.5   &       28.6  &       28.0 &  $       -31^{+   64}_{-   35}$  &  $<$ 0.050  &  $<$ 0.027\\ 
J150332.93+440120.7   &  2.040   &   3.565   &   $       5815 $&       8083  &        13.5   &       12.5  &       12.2 &  $      -218^{+  265}_{-  199}$  &  $<$ 0.236  &  $<$ 0.177\\ 
J235702.54$-$004824.0   &  2.994   &   3.543   &   $        -36 $&       1479  &         8.3   &        7.2  &        7.4 &  $        72^{+  197}_{-  267}$  &  $<$ 0.176  &  $<$ 0.239\\ 
J235859.47$-$002426.2   &  1.759   &   5.132   &  $        3416 $&       5609  &        16.4   &       28.1  &       31.8 &  $        -1^{+   99}_{-  159}$  &  $<$ 0.061  &  $<$ 0.098\\
\label{tbl:bal_params}     
\end{longtable}     
\tablecomments{
(a) Redshifts are from \cite{Hewett10}. 
(b) $\Delta t$ is calculated in the quasar rest-frame between the earliest (SDSS) and latest (TDSS) epochs.  
(c) For full-trough upper limits, $v_{\rm min}$ and $v_{\rm max}$ given are for the entire BAL; for the partial-trough upper limits, $v_{\rm min}$ and $v_{\rm max}$ listed are the actual trough limits used when computing the upper limits (see Section \ref{sec:upperlimits}).
(d) EW 1 refers to measurements from Epoch 1 (SDSS), EW 2 refers to Epoch 2 (BOSS) and EW 3 refers to Epoch 3 (TDSS). EW measurements were made using the individual $v_{\rm min}$ and $v_{\rm max}$ measured in each epoch, rather than the adopted $v_{\rm min}$ and $v_{\rm max}$ that spans the entire BAL in all three epochs as described in Section \ref{sec:troughs}. In cases where we had multiple BAL troughs in some epochs that we consider a part of a single complex, we show the sum of the EW measurements for all sub-troughs in a complex. Regardless of whether the upper limits were measured using the full-trough or just part of it, the EW values listed include the entire BAL. 
(e) Velocities are given in units of \kms and velocity shift uncertainties are 3$\sigma$ uncertainties rather than the standard 1$\sigma$ uncertainties. This is done because the acceleration upper limits were all calculated with the 3$\sigma$ upper limits rather than the 1$\sigma$ upper limits. 
(f) The measured velocity shift and acceleration upper limits are measured between the earliest (SDSS) and latest (TDSS) epochs.    
(g) All upper limits are quoted at 3$\sigma$ significance. } 
\setlength{\tabcolsep}{6pt} 
\end{center}   

\end{document}